\documentclass[
  10pt,
  twocolumn,
  a4paper,
  pra,
  aps,
  amssymb,
  amsmath,
  superscriptaddress,
  amsfonts,
  nofootinbib,
  eprint,
]{revtex4-1}

\usepackage[noEnd=true,indLines=true,spaceRequire=false,rightComments=false]{algpseudocodex}
\usepackage{graphicx} 
\usepackage[normalem]{ulem}
\usepackage{xcolor}
\usepackage{physics}
\usepackage{acronym}
\usepackage{mathtools}
\usepackage{amsmath}
\usepackage{amsthm}
\usepackage{amsfonts}
\usepackage{natbib}
\usepackage{appendix}
\usepackage{quantikz}
\usepackage{graphicx}
\usepackage{subcaption}
\usepackage{multirow}
\usepackage[dvipsnames]{xcolor}
\usepackage{hyperref}
\usepackage[capitalize]{cleveref}

\let\vec\boldsymbol
\let\Vec\boldsymbol
\newcommand{\vtheta}{\Vec{\theta}_t}

\newcommand{\tvtheta}{\Tilde{\Vec{\theta}}}
\newcommand{\parvtheta}{\left(\Vec{\theta}_t\right)}
\newcommand{\dvtheta}{\dot{\Vec{\theta}}_t}
\newcommand{\ttheta}{\Tilde{\Vec{\theta}}}
\newcommand{\init}{\psi_{o}}

\newcommand{\ta}{\ket{{\alpha}}}
\newcommand{\tb}[1]{\ket{{\beta}_{#1}}}
\newcommand{\tc}[1]{\ket{{\gamma}_{#1}}}

\newcommand{\varpsi}{\psi\left(\vtheta\right)}
\newcommand{\parvarpsi}[1]{\partial_{#1}\psi\left(\vtheta \right)}

\newcommand{\U}{U\left(\vtheta\right)}
\newcommand{\Ud}{U^\dag\left(\vtheta\right)}
\newcommand{\pU}[1]{\partial_{#1}\U}

\newcommand{\spsapert}{\boldsymbol{\delta}}
\DeclareMathOperator*{\argmin}{argmin}

\DeclareMathOperator*{\Var}{Var}

\mathtoolsset{showonlyrefs=false}

\theoremstyle{plain}

\theoremstyle{definition}
\newtheorem{definition}{Definition}

\definecolor{pallette1}{HTML}{9EACFF}
\definecolor{pallette2}{HTML}{B69CDF}
\definecolor{pallette3}{HTML}{C68CBF}
\definecolor{pallette4}{HTML}{D17BA0}
\definecolor{pallette5}{HTML}{D4C065}
\definecolor{pallette6}{HTML}{BDC576}
\definecolor{pallette7}{HTML}{A2C987}
\definecolor{pallette8}{HTML}{82CC97}

\newcommand{\roundparamgate}[2]{\gate[2,style={draw,line width=0.5pt,fill=#1,rounded corners=2pt,inner sep=1pt,transform shape,scale=0.75}]{#2}}

\newcommand{\shadedparamgate}[1]{\gate[2,style={draw,line width=0.5pt,rounded corners=2pt,inner sep=1pt,transform shape,scale=0.75,dashed}]{#1}}

\newcounter{algorithm}
\renewcommand{\thealgorithm}{\arabic{algorithm}}

\newcommand{\algorithmcaption}[1]{
  \refstepcounter{algorithm}
  \captionof*{figure}{\textbf{Algorithm \thealgorithm:} #1}
}

\newacro{SPSA}{Simultaneous Perturbation Stochastic Approximation}
\newacro{VarQTE}{Variational Quantum Time Evolution}
\newacro{VarQITE}{Variational Quantum Imaginary Time Evolution}
\newacro{VarQRTE}{Variational Quantum Real Time Evolution}
\newacro{QGT}{Quantum Geometric Tensor}
\newacro{ODE}{Ordinary Differential Equation}
\newacro{SKQD}{Sample Based Krylov Quantum Diagonalization}
\newacro{AQC}{Approximate Quantum Compiling}
\newacro{MPS}{Matrix Product State}
\newacroplural{MPS}[MPSs]{Matrix Product States}
\newacro{FLOPs}{Floating Point Operations}
\newacro{FLOP}{Floating Point Operation}
\newacro{HVA}{Hamiltonian Variational Ansatz}
\newacro{SLE}{System of Linear Equations}
\newacro{VC}{Variational Compression}

\begin{document}

\title{Bowtie VarQTE: A Resource-Efficient Quantum State Preparation Primitive}
\author{Marc Drudis}
\affiliation{IBM Quantum, IBM Research Europe — Zurich, Rüschlikon 8803, Switzerland}
\author{Alberto Baiardi}
\affiliation{IBM Quantum, IBM Research Europe — Zurich, Rüschlikon 8803, Switzerland}
\author{Mattia Chiurco}
\affiliation{IBM Quantum, IBM Research Europe — Zurich, Rüschlikon 8803, Switzerland}
\affiliation{Institute for Theoretical Physics, ETH Zurich, Switzerland}
\author{Francesco Tacchino}
\affiliation{IBM Quantum, IBM Research Europe — Zurich, Rüschlikon 8803, Switzerland}
\author{Stefan Woerner}
\affiliation{IBM Quantum, IBM Research Europe — Zurich, Rüschlikon 8803, Switzerland}
\author{Christa Zoufal}
\affiliation{IBM Quantum, IBM Research Europe — Zurich, Rüschlikon 8803, Switzerland}

\date{\today}

\begin{abstract}

The preparation of quantum states is a fundamental requirement for many quantum algorithms. 
A native route to preparing physically structured states is based on short-time simulation of dynamical processes, such as real or imaginary time evolution.
This work presents a resource-efficient framework for the approximation thereof with \textit{bowtie \ac{VarQTE}} which uses classical simulation where possible and quantum resources where necessary.
We introduce a framework that leverages existing causal light-cones to minimize quantum resource requirements in the evaluation of gradient and quantum geometric tensor terms by utilizing classical simulation methods for causally relevant subcircuits. This in turn enables exact parameter updates according to  McLachlan's variational principle and, thereby, improves numerical stability. 
We conduct a comparison with a state preparation method that is based on a tensor-network compiled Trotter algorithm: approximate quantum compilation (AQC). In recent work, this approach has shown impressive performance. However, its key-bottleneck is the necessity to have a classical (approximate) representation of the target state. 
Our numerical experiments indicate that bowtie VarQTE can achieve comparable fidelities without this requirement.
We further illustrate how bowtie VarQTE can facilitate a state-preparation pipeline that combines the simulation of imaginary and real time evolution for a sample-based quantum algorithm.
In fact, results on 2D systems show how bowtie VarQTE can reduce the quantum requirements compared to standard, sample-based Krylov diagonalization calculations.
 Our results indicate that VarQTE is a promising primitive for the preparation of physically structured quantum states that reduces requirements on quantum resources by leveraging existing structures and the associated possibility of enabling classical simulations.
\end{abstract}

\maketitle

\section{Introduction}
Efficient quantum state preparation underpins the performance of a wide range of quantum algorithms, playing a decisive role in applications from ground state search algorithms, such as sampling-based quantum diagonalization and quantum phase estimation to Hamiltonian simulation, and even partial differential equation simulations. In many such algorithms, the success probability, accuracy, and overall resource requirements are critically dependent on the quality of the initial state, often relying on substantial overlap with specific eigenstates or low-energy subspaces of interest. Consequently, the cost of preparing these structured input states can dominate the total computational complexity, even hinder potential quantum speedups. Hence, designing scalable, robust, and hardware-efficient state preparation protocols has emerged as a central obstacle in realizing the practical potential of quantum algorithms across diverse application domains.

Real- and imaginary-time evolution provide a natural framework for the preparation of \textit{physically motivated quantum states} by driving a quantum system toward a desired target-state manifold. 
The prevailing state-of-the-art approach to simulating the time evolution of a $k$-local Hamiltonian is based on product-formula methods constructed via Trotterization \cite{trotter1959, suzuki1976, lloyd1996}. For real-time evolution, such Trotter circuits are straightforward to implement and are accompanied by rigorous analytical bounds that quantify the trade-off between simulation error and circuit depth. Nevertheless, attaining high-precision evolution typically requires circuit depths that are prohibitively large for early-generation quantum hardware. The situation is further complicated for imaginary-time evolution, where Trotter-based implementations do not directly yield unitary dynamics and instead rely on mappings to effective unitary ansätze \cite{motta_determining_2020}. Closely related are adiabatic state-preparation protocols, which offer a conceptually simple route to preparing correlated ground states, but are fundamentally limited by the minimum spectral gap encountered along the interpolation path, leading to long evolution times and unfavorable scaling for many systems of practical interest.

Variational Quantum Time Evolution (\ac{VarQTE}) \cite{yuan_theory_2019, McArdleVarQITE19} offers an alternative to Trotter and adiabatic simulation that is directly applicable to real and imaginary time evolution.
It employs McLachlan's variational principle \cite{McLachlan64} to propagate the parameters of a variational ansatz over discretized time steps with ordinary differential equation (ODE) solvers. The goal is to project the evolution trajectory onto the variational manifold given by a parameterized ansatz.
Unlike Trotter approaches that approximate the full evolution operator, \ac{VarQTE} focus on approximating the action of the time evolution on a specific initial state.
Notably, the underlying variational principle facilitates the efficient computation of a-posteriori state preparation error bounds \cite{Local-in-TimeErrorMartinazzo20, Zoufal_2023ErrorBounds}--an important contrast to most variational quantum algorithms.
The feasibility of the practical realization and accuracy of \ac{VarQTE} strongly depend on the required number of circuit evaluations, the expressivity of the variational ansatz, and the chosen time integration method.

The main contributions of this work are as follows: First, we introduce a method to compute the \ac{QGT} of a variational quantum circuit at system sizes that lie beyond the reach of existing approaches--which are discussed in Appendix~\ref{app:alternative_qgt_strategies}. 
Our approach builds upon the mathematical structure of the \ac{QGT}. 
More specifically, we identify causal light-cones \cite{eddins_lightcone_2024} and effectively isolate the relevant sub-systems such that the problem is decomposed into the independent simulation of smaller, tractable sub-circuits which can be analytically computed.
This capability, in turn, enables a systematic study of \ac{VarQTE} at scale, based on causal locality: \textit{bowtie \ac{VarQTE}}.
The underlying framework allows the balanced use of classical and quantum resources.
More specifically, supporting \ac{VarQTE} with classical simulation techniques--where possible--enables approximations with good fidelities. 
The use of quantum resources can, thus, be focused on classically difficult operations such as sampling from states with high degrees of entanglement or magic.
This approach ensures numerical stability and physical reliability throughout the evolution while offering computational efficiency. For the system sizes considered in this work, alternative approaches are either prohibitively slow or yield only approximate estimates.
Finally, we show that bowtie \ac{VarQTE} can be used for state-preparation by combining imaginary- and real-time \ac{VarQTE} within a sampling-based quantum algorithm.
We benchmark our approach against \ac{AQC} \cite{Robertson2025_AQC, pennington2026preparing100qubitsymmetryprotectedtopological, rscApproximateQuantum}, a widely used state-preparation method that employs classical tensor-network techniques \cite{Deutsch89TensorNetworks} to compile Trotter-based quantum circuits into short-depth approximations. Notably, this method--unlike bowtie VarQTE--requires access to a classical (approximation) of the target system. We find that our method achieves similar state preparation fidelities by directly following the variational principle underlying McLachlan~\cite{McLachlan64}.
Furthermore, we explore a state-preparation protocol of a 2D system in a sampling-based quantum diagonalization algorithm which requires the combination of both real- and imaginary-time evolution. 
Finally, we analyze the computational resources required to evaluate the quantities governing bowtie VarQTE and compare it to \ac{AQC}.

The remainder of this paper is structured as follows. Firstly, a brief overview of VarQTE is given in Sec.~\ref{sec:VarQTE}. We introduce bowtie VarQTE in Sec.~\ref{sec:methods} and present the main results of this work in Sec.~\ref{sec:results}. Finally, we provide a discussion and an outlook in Sec.~\ref{sec:discussion}.

\section{Variational Quantum Time Evolution}
\label{sec:VarQTE}

\subsection{McLachlan's Variational Principle}

A central challenge in quantum simulation is the preparation of states that accurately approximate the dynamics of a target Hamiltonian underlying real or imaginary time evolution.
 \ac{VarQTE} enables an approximate representation of such time evolved states.
\begin{definition}[Variational Quantum Time Evolution]\label{def:varqte} 
\ac{VarQTE} describes the approximation of the real or imaginary time evolution
\begin{align}
    \partial_t \ket{\psi^{\text{real}}(t)} &= -iH \ket{\psi^{\text{real}}(t)}, \\
\partial_{\tau} \ket{\psi^{\text{imag}}(\tau)} &= -\frac{H \ket{\psi^{\text{imag}}(\tau)}}{\bra{\psi^{\text{imag}}(\tau)}H \ket{\psi^{\text{imag}}(\tau)}},
\end{align}
of a state $\ket{\psi(t)}$ according to a $k-$local Hamiltonian $H=\sum_{i=0}^{N_h} c_i h_i $ for time $t$ with an $n$-qubit parameterized  ansatz $\lvert{\psi(\vtheta)}\rangle=U(\vtheta)\ket{0}^{\otimes n}$ where $\vtheta\in \mathbb{R} ^ m$ by mapping the time dependence onto the ansatz parameters: $\boldsymbol{\theta} \rightarrow \vtheta $. 
The parameter propagation follows a time-discretized \ac{ODE} defined through McLachlan's variational principle.
This principle requires that at each time step the following \textit{gradient} quantities are computed for real and imaginary time, respectively:
\begin{align}
  b^{\text{real}}_i(\vtheta) &= \Im\left( \langle \partial_i \psi (\vtheta)| H | \psi (\vtheta)\rangle \right. - \\
  &\left. \hspace{11mm} \langle \partial_i \psi (\vtheta)| \psi (\vtheta)\rangle E_{\vtheta}\right), \label{eq:breal} \\
  b^{\text{imag}}_i(\vtheta) &= -\Re\left( \langle \partial_i \psi(\vtheta) | H | \psi(\vtheta) \rangle\right), \label{eq:bimag}
\end{align}
with $E_{\vtheta} = \langle \psi(\vtheta) | H | \psi(\vtheta) \rangle $
 and the \ac{QGT}
\begin{align}
    \label{eq:qgt}
  \begin{split}
    g_{ij}(\vtheta) = &\Re\left( \langle \partial_i \psi (\vtheta)| \partial_j \psi (\vtheta)\rangle \right.\\
    &\left. - \langle \partial_i \psi (\vtheta)| \psi (\vtheta)\rangle \langle \psi(\vtheta) | \partial_j \psi(\vtheta) \rangle \right),
  \end{split}
\end{align}
where \( |\partial_i \psi(\vtheta)\rangle = \frac{\partial|\psi(\vtheta)\rangle}{\partial \theta_i} \).
This in turn defines the following \ac{SLE}
\begin{align}
  \label{eq:sle}
   g(\vtheta) \dvtheta = \vec{b}(\vtheta).
\end{align}
Solving the equation above for $\dvtheta=\frac{\partial \vtheta}{\partial_t}$ allows us to propagate the ansatz parameters using an ODE solver such that the evolution approximates the actual time evolution.
\end{definition}

The number of gradient entries scales linearly with the number of parameters $N_p$, while the number of \ac{QGT} entries scales quadratically.
For non-trivial system sizes, this alone can already represent a significant resource challenge.
Furthermore, the underlying \ac{SLE} given in~\cref{eq:sle} is typically ill-conditioned.
As a consequence, small errors in the QGT can result in large errors in the parameter update.
These errors, in turn, can lead to unstable and nonphysical dynamics in \ac{VarQTE}~\cite{DualQITE}.
Therefore, each QGT entry must be evaluated with high numerical precision.
This requirement implies a large number of shots on the quantum computer.
Consequently, the quadratic scaling of the \ac{QGT} is widely identified in the literature as the dominant computational bottleneck in \ac{VarQTE}.
Achieving the necessary accuracy for an effective \ac{VarQTE} implementation with these methods remains prohibitively expensive in practice.
We provide an overview over the most established approximation methods in the next section and discuss their bottlenecks in more detail.

\section{Methods}
\label{sec:methods}

Next, we discuss our framework to approach the two main bottlenecks of \ac{VarQTE}: the large number of required quantum measurements and the numerical instability of the underlying \ac{SLE}. More specifically, we introduce a method that, when possible, computes gradient and \ac{QGT} entries exactly using classical resources. This approach improves algorithmic stability and reduces the quantum workload.

\subsection{Bowtie Gradient and QGT Computation}\label{sec:bowtiegradientandqgtcomputation}

In the following, we consider a variational ansatz as defined below.
 \begin{definition}[Ansatz]\label{def:circuit}
Let our ansatz be a parameterized circuits $U(\vtheta)$ of the form
$$
U(\boldsymbol{\theta}) = \prod_{k=0}^{N_p-1} C_k R_{P_k}(\theta_k) \,,
$$
where $C_k$ are Clifford gates and $R_{P_k}(\theta_k)=e^{-i P_k \theta_k}$, with $P_k$ a Pauli operator.
We initialize our time dependent parameters $\vtheta\in \mathbb{R} ^{N_k}$ such that:
\begin{equation}
    U(\boldsymbol{\theta}^0)\ket{0}^{\otimes n} = \ket{\psi(t=0)}.
\end{equation}
\end{definition} 
 To compute the elements in the gradient $\vec{b}(\vtheta)$ given in~\cref{eq:breal,eq:bimag}, as well as the QGT $g(\vtheta)$ given in~\cref{eq:qgt}, we may exploit the underlying light-cone structure--see Definition~\ref{def:light_cones}. In particular, we show how classical computational resources can be leveraged to evaluate these quantities precisely for systems that exhibit sufficient structure. 
\begin{definition}[Light-cones]\label{def:light_cones} Provided any fixed $N_p$-parameter circuit $U(\vtheta)$ from Definition \ref{def:circuit}, the \textit{light-cone} 
$\Delta_{P}^{U}$
is defined as the number of qubits on which  $U^\dagger(\vtheta)P U(\vtheta)$ acts non-trivially with respect to a Pauli observable $P$ for any $\vtheta \in [-\pi, \pi]^m$.
\end{definition}

\begin{figure}[ht]
  \centering
    \begin{subfigure}[b]{0.95\linewidth}
  \resizebox{\linewidth}{!}{
    \begin{quantikz}[
        row sep={0.45cm,between origins},
        column sep=0.38cm
      ]
      \qw 
      & \roundparamgate{pallette8}{\theta_{00}} &  \qw
      & \roundparamgate{pallette5}{\theta_{10}} &  \qw
      & \roundparamgate{pallette8}{\theta_{20}} &  \qw 
      &\qw \\
      \qw 
      &\qw & \roundparamgate{pallette8}{\theta_{05}} 
      &\qw & \roundparamgate{pallette5}{\theta_{15}} 
      &\qw & \roundparamgate{pallette1}{\theta_{25}}
      &\qw \\
      \qw 
      & \roundparamgate{pallette1}{\theta_{01}} &  \qw
      & \roundparamgate{pallette8}{\theta_{11}} &  \qw
      & \roundparamgate{pallette5}{\theta_{21}} &  \qw 
      &\qw \\
      \qw 
      &\qw & \roundparamgate{pallette1}{\theta_{06}} 
      &\qw & \roundparamgate{pallette1}{\theta_{16}} 
      &\qw & \roundparamgate{pallette5}{\theta_{26}}
      &\qw \\
      \qw 
      & \roundparamgate{pallette5}{\theta_{02}} &  \qw
      & \gate[2,style={draw,line width=1.5pt,fill=pallette4,inner sep=1pt,transform shape,scale=0.75,}]{\theta_{12}} & \qw
      & \roundparamgate{pallette1}{\theta_{22}} &  \qw 
      &\qw \\
      \qw 
      &\qw & \roundparamgate{pallette5}{\theta_{07}} 
      &\qw & \roundparamgate{pallette8}{\theta_{17}} 
      &\qw & \roundparamgate{pallette8}{\theta_{27}}
      &\qw \\
      \qw 
      & \roundparamgate{pallette1}{\theta_{03}} &  \qw
      & \roundparamgate{pallette1}{\theta_{13}} &  \qw
      & \roundparamgate{pallette5}{\theta_{23}} &  \qw 
      &\qw \\
      \qw 
      &\qw & \roundparamgate{pallette8}{\theta_{08}} 
      &\qw & \roundparamgate{pallette8}{\theta_{18}} 
      &\qw & \roundparamgate{pallette1}{\theta_{28}}
      &\qw \\
      \qw 
      & \roundparamgate{pallette5}{\theta_{04}} &  \qw
      & \roundparamgate{pallette5}{\theta_{14}} &  \qw
      & \roundparamgate{pallette1}{\theta_{24}} &  \qw 
      &\qw \\
      \qw 
      &\qw & \roundparamgate{pallette8}{\theta_{09}} 
      &\qw & \roundparamgate{pallette1}{\theta_{19}} 
      &\qw & \roundparamgate{pallette8}{\theta_{29}}
      &\qw \\
      \qw 
      & \qw & \qw 
      & \qw & \qw 
      & \qw & \qw
      & \qw 
    \end{quantikz}
  }
  \subcaption{Initial circuit with the selected rotation gate $R_P(\theta_{12}) = e^{-iP\theta_{12}} $.}
  \end{subfigure}
    \begin{subfigure}[b]{0.95\linewidth}
    \label{subfig:bowtie}
  \resizebox{\linewidth}{!}{
    \begin{quantikz}[row sep={0.45cm,between origins},column sep=0.38cm]
      \qw 
      & \shadedparamgate{+\theta_{00}} &  \qw
      & \shadedparamgate{+\theta_{10}} 
      & \shadedparamgate{-\theta_{10}} 
      & \qw 
      & \shadedparamgate{-\theta_{00}}
      &\qw \\
      \qw 
      &\qw & \shadedparamgate{+\theta_{05}} 
      &\qw
      &\qw & \shadedparamgate{-\theta_{05}} 
      &\qw 
      &\qw \\
      \qw 
      & \roundparamgate{pallette1}{+\theta_{01}} &  \qw
      & \shadedparamgate{+\theta_{11}} 
      & \shadedparamgate{-\theta_{11}} 
      & \qw 
      & \roundparamgate{pallette1}{-\theta_{01}} 
      &\qw \\
      \qw 
      &\qw & \roundparamgate{pallette1}{+\theta_{06}} 
      &\qw 
      &\qw & \roundparamgate{pallette1}{-\theta_{06}} 
      &\qw \\
      \qw 
      & \roundparamgate{pallette5}{+\theta_{02}} 
      &  \qw
      & \gate[2,style={draw,line width=1.5pt,xshift=2.0em,fill=pallette4,inner sep=1pt,transform shape,scale=0.75,minimum width = 7em}]{P} \qw & \qw
      & \qw & \roundparamgate{pallette5}{-\theta_{02}} 
      &\qw \\
      \qw 
      &\qw & \roundparamgate{pallette5}{+\theta_{07}} 
      &\qw
      &\qw & \roundparamgate{pallette5}{-\theta_{07}} 
      &\qw
      &\qw \\
      \qw 
      & \roundparamgate{pallette1}{+\theta_{03}} &  \qw
      & \shadedparamgate{+\theta_{13}} 
      & \shadedparamgate{-\theta_{13}} 
      & \qw & \roundparamgate{pallette1}{-\theta_{03}} 
      &\qw \\
      \qw 
      &\qw & \shadedparamgate{+\theta_{08}} 
      &\qw 
      &\qw & \shadedparamgate{-\theta_{08}} 
      &\qw 
      &\qw \\
      \qw 
      & \shadedparamgate{+\theta_{04}} &  \qw
      & \shadedparamgate{+\theta_{14}} 
      & \shadedparamgate{-\theta_{14}} 
      & \qw & \shadedparamgate{-\theta_{04}} 
      &\qw \\
      \qw 
      &\qw & \shadedparamgate{+\theta_{09}} 
      &\qw 
      &\qw & \shadedparamgate{-\theta_{09}} 
      &\qw 
      &\qw \\
      \qw 
      & \qw & \qw & \qw 
      & \qw & \qw & \qw 
      &\qw \\
    \end{quantikz}
    }
  \subcaption{Bowtie circuit associated with $ \theta_{12} $. First, the gate is replaced by its Pauli generator $P$ and then $ U^\dagger $ is appended. As a result, all gates outside the lightcone cancel out. 
  As per our definition, this circuit is the \textbf{bowtie} $B_{\beta}^{12}$, where $\beta$ indicates that it is the bowtie of a parameter and $12$ is the index of such parameter. The state generated by the action of such circuit on $\ket{0}$ would be the \textbf{bowtie state} $\ket{\beta_{12}}$.
 } 
  \end{subfigure}
  \caption{Construction of a bowtie circuit for gradient extraction of a variational parameter.}
  \label{fig:bowtie_illustration}
\end{figure}
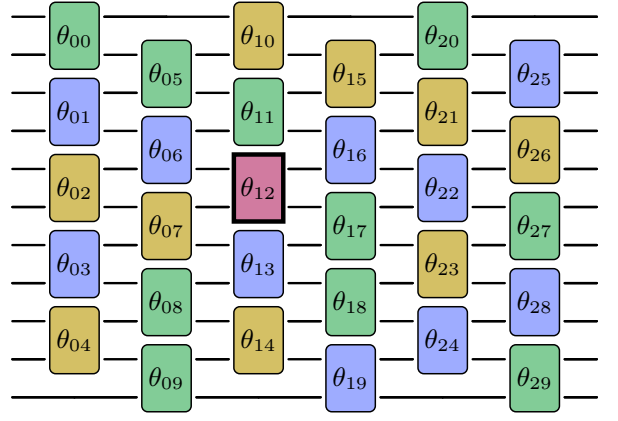
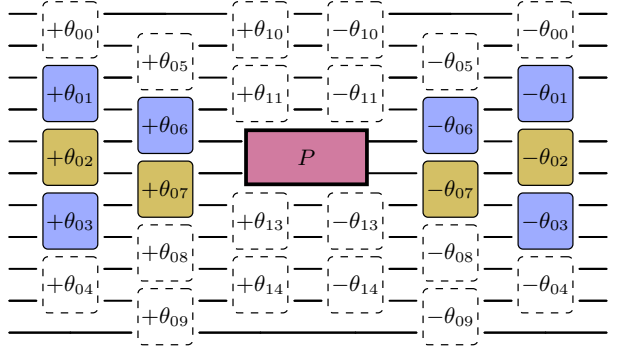

We can observe that the elements in $\vec{b}(\vtheta)$ and $g(\vtheta)$ are based on complex-valued inner products. These inner products are invariant under unitary transformations such that we can rewrite them as:

\begin{align}
\label{eq:overlaps1}
   \langle \partial_i \psi (\vtheta)| h_j | \psi (\vtheta)\rangle &=
    \braket{{\beta}_i}{{\gamma}_j} ,\\
    \label{eq:overlaps2}
    \langle \partial_i \psi (\vtheta)| \psi (\vtheta)\rangle&=\braket{{\beta}_i}{{\alpha}} ,\\
    \label{eq:overlaps3}
    \langle \psi(\vtheta) | h_j | \psi(\vtheta) \rangle &= \braket{{\alpha}}{{\gamma}_j}  ,\\
    \label{eq:overlaps4}
    \langle \partial_i \psi (\vtheta)| \partial_j \psi (\vtheta)\rangle &= \braket{{\beta}_i}{{\beta}_j},
\end{align}
where
\begin{align}
\label{eq:alpha}
  \ta &\coloneqq \Ud \ket{\varpsi}= \ket{0},\\
  \label{eq:beta}
  \tb{i} &\coloneqq \Ud \ket{\parvarpsi{i}} = B_{\beta}^p \parvtheta \ket{0},\\
  \label{eq:gamma}
  \tc{j} &\coloneqq \Ud  h_j\ket{\varpsi }= B_{\gamma}^j \parvtheta \ket{0},
\end{align}
with $B_{\beta}^i \parvtheta \coloneqq\Ud \pU{i}$ for $i\in\left\{0, \ldots, N_p-1\right\}$ and $B_{\gamma}^j \parvtheta \coloneqq \Ud h_j \U $ for $j\in\left\{0, \ldots, N_h-1\right\}$.
The state $\ta$ is simply a product state, while $\tb{i}$ and $\tc{j}$ may be simplified according to their underlying symmetric light-cone structure. For $\tc{j}$ it is straightforward to see that $B_{\gamma}^q \left(\vtheta\right)$ induces a light-cone $\Delta_{h_q}^{U}$ as per Definition~\ref{def:light_cones}.
To find the light-cone structure underlying $\tb{i}$, we use the fact that the gradients of the parameterized gates within our ansatz, see Definition~\ref{def:circuit}, correspond to $\frac{\partial e^{-iP\theta}}{\partial \theta}=-iPe^{-iP\theta}$. Hence, we can rewrite 
\begin{equation}
B_{\beta}^i \left(\vtheta\right)= W^{\dagger}V^{\dagger}VP_i W = W^{\dagger}P_i W, 
\end{equation}
for unitaries $W$ and $V$ such that the respective light-cone structure corresponds to $\Delta_{P_j}^{W}$.
Inspired by the underlying structure--illustrated in \cref{fig:bowtie_illustration}--we shall from now on refer to $\tb{i}$ and $\tc{j}$ as \textbf{bowtie states} and the underlying light-cone inducing unitaries $B_{\beta}^i$ and $B_{\gamma}^j$ as \textbf{bowties}.
Details on the efficient computation of the overlaps are given in Appendix~\ref{app:overlap_qubit_intersect}.

\subsection{Algorithm}
\begin{figure}[ht!]
  \centering
  \begin{algorithmic}
    \algorithmcaption{Bowtie-\ac{VarQTE}}
    \label{algo:Bowtie}
    \Require qc: ansatz as per Definition~\ref{def:circuit} with $\vtheta\in\mathbb{R}^{N_p}$
    \Require $H = \sum\limits^{N_h}_q c_q h_q$: with $c_q\in\mathbb{R}$ and $h_q$ k-local Paulis
    \State \Comment{Pre-compute}
    \For i in $\{0, \ldots, N_p-1\}$
    \State Construct $B_{\beta}^{i} $
    \State Store indices of qubits in $\Delta_{P_i}^W$
    \EndFor
    \For j in $\{0, \ldots, N_h-1\}$
    \State Construct $B_{\gamma}^i$
    \State Store indices of qubits in $\Delta_{h_i}^U$
    \EndFor
    \For i in $\{0, \ldots, N_p-1\}$
    \For  j in $\{0, \ldots, N_p-1\}$
    \State Pre-compute $\braket{{\beta}_p}{{\beta}_j}$ contraction
    \EndFor
    \For j in $\{0, \ldots, N_h-1\}$
    \State Pre-compute $\braket{{\beta}_p}{{\gamma}_j}$ contraction
    \EndFor
    \EndFor
    \State \Comment{Time evolution}
    \While{$t < t_{\text{final}}$}
    \State Bind parameters $\vtheta$
    \State Simulate all \textit{bowtie states}
    \State Evaluate $\Vec{b}\parvtheta$
    \State Evaluate $g\parvtheta$
    \State Solve SLE given in~\cref{eq:sle}
    \State Propagate parameters with ODE solver
    \EndWhile
    \Return $\vec{\theta}^{t_\text{final}}$
  \end{algorithmic}
\end{figure}

The full bowtie VarQTE algorithm is given in pseudocode in Algorithm~\ref{algo:Bowtie} and a reference code implementation can be found in \cite{zenododrudis}.
The central idea is to separate the workflow into two stages: a one-time pre-computation step that depends on the light-cone structure of the bowties, and a per-timestep simulation step.
The former stores intermediate results, while the latter is dominated by computation of the bowtie states for $\vtheta$, enabling the most efficient execution.

At first, we pre-compute the bowties independent of the numerical values of $\vtheta$. 
\begin{enumerate}
  \item \textbf{Bowtie templates.} We construct and store all bowties $B_{\beta}^p $ and $B_{\gamma}^q$ in a parameterized form.
  \item \textbf{Index tracking light-cone qubits.} For each bowtie, we track which qubits are within the light-cone. 
  \item \textbf{Pre-computed overlaps.} For each pair of \textit{bowtie states} we keep track of what indices of each tensor need to be contracted and in which order.
\end{enumerate}
For each time-step in the \ac{VarQTE} propagation, we can then bind the parameters $\vtheta$ to the pre-computed overlaps which allow us to compute $\dot{\vtheta}$ by solving~\cref{eq:sle}, and propagate the parameters accordingly.
Notably, most operations are easy to parallelize: the bowties can be simulated independently, and the overlaps can be evaluated independently with the corresponding tensor-representation of the bowtie states. 
In fact, the engineering details have a strong impact on how efficient the algorithm may be executed. 

The main factors that enable an efficient practical execution of bowtie-\ac{VarQTE} are the ability to use (i) bowtie state pre-computation, (ii) overlap computations that are limited to the qubits involved in the overlapping light-cones, and (iii) parallel execution. 
Additionally, we note a close structural analogy between bowties arising from parameterized gates and those induced by Hamiltonian terms, as both give rise to overlaps of states generated by local circuit extensions; this observation motivates a unified treatment in terms of a generalized \ac{QGT}, discussed in Appendix~\ref{appendix:aqc2mclachlan}.

\section{Results}
\label{sec:results}

This section presents the main results and describes the key strengths and limitations of our approach.
Firstly, we investigate the expressivity of bowtie VarQTE by comparing it to Trotter- and an AQC-based \cite{Robertson2025_AQC, jaderberg2025variationalpreparationnormalmatrix} state preparation methods in Sec.~\ref{sec:expressivity}.
Secondly, we demonstrate a workflow that integrates real and imaginary bowtie \ac{VarQTE} for state preparation in a quantum sampling diagonalization methods in Sec.~\ref{sec:SKQD}.
Finally, Sec.~\ref{sec:comp_cost} provides a resource analysis.

\subsection{Expressivity Analysis}
\label{sec:expressivity}

This section presents a comparison of how effectively different methods encode time-evolved quantum states.
More specifically, we compare bowtie \ac{VarQTE} to Trotterized circuits and to \ac{AQC}, a state-of-the-art state preparation approach.
The latter corresponds to a tensor-network based compilation method \cite{Sharma_2020, Jones2022robustquantum, jaderberg2025variationalpreparationnormalmatrix, Robertson2025_AQC},
which can achieve significant reductions in circuit depth when encoding time-evolved states compared to standard Trotterized approaches.
However, since \ac{AQC} relies on classical tensor-network simulation, it is strongly dependent on the bond-dimension of the underlying problem and it requires access to a simulation of the target state.
Conversely, the \ac{VarQRTE} framework is primarily constrained by the sizes of the relevant light-cones induced by the ansatz and the Hamiltonian and may, thus, be better suited to 2D or higher-dimensional topologies. 

\subsubsection{Approximate Quantum Compilation}
The core idea behind \ac{AQC} is to simulate the unitary evolution using an \ac{MPS} which is then compressed it into a parameterized circuit of fixed depth. 
Mathematically, this involves approximating the exact evolution operator $U(t)=e^{-i H t}$
by a variational ansatz, e.g., of the form given in Definition~\ref{def:circuit}.

Independently of the choice of the cost function, its optimization requires that a classical representation of the targe state $e^{-i H t} \ket{\init}$ is available.
An exact representation of this state is only feasible for relatively small qubit numbers and for utility-scale simulations one must resort to either approximations or restraining the approach to subsets of the full circuit.

The fact that \ac{AQC} relies on the ability to simulate (or at least approximate) the unitary evolution classically, provides  an important differentiator to \ac{VarQTE}.
In \ac{AQC}, such an approximate representation is obtained with tensor networks and, specifically, with \acp{MPS}~\cite{Schollwock2011_DMRG}.
\ac{MPS} is a variational ansatz that can accurately represent the ground-state of one-dimensional quantum systems with resources growing polynomially with the system size, as opposed to the exponential growth needed to represent generic quantum states.
Based on this, \ac{AQC} optimizes the cost function given above in two steps.
First, the \ac{MPS} representation $\ket{\psi_\text{\ac{MPS}}}$ of $e^{-i H t} \ket{\init}$ is obtained with the time-evolving block-decimation algorithm~\cite{Vidal2003_TEBD} -- an algorithm that enables time-evolving a wave function by retaining its \ac{MPS} structure.
The resulting \ac{MPS} is then compiled into a quantum circuit.
Although exact compilation strategies for \ac{MPS} wave functions exist~\cite{Schoen2005_MPS-Preparation}, the resulting quantum circuit becomes deep for large bond dimensions.
Conversely, AQC adopts a variational compilation strategy: the $e^{-i H t} \ket{\init}$ is encoded through a parametrized circuit $V(\theta_i) \vert 0 \rangle$, where the $\theta_i$ parameters are optimized \emph{classically} by maximizing the overlap 
Naturally, the accuracy of the resulting compilation depends on the form of the parametrized circuit $V(\theta_i)$.
Since existing \ac{AQC} implementations are limited to \ac{MPS}-based classical simulations, they are mainly suitable for 1D systems. Nevertheless, extensions of \ac{AQC} to tensor networks which are better suited to higher-dimensional systems are possible and may provide interesting compilation possibilities in the future. 

\subsubsection{Comparison}\label{sec:comparaison}
We compare bowtie \ac{VarQRTE} with a second-order Suzuki--Trotter baseline and an \ac{AQC}-compiled version thereof on one-dimensional Heisenberg chains.
The simulated model is the nearest-neighbor Heisenberg Hamiltonian on a 1D lattice,
\begin{equation}
H_{\mathrm{1D}} = \sum_{i=1}^{N-1} \left(X_iX_{i+1}+Y_iY_{i+1}+Z_iZ_{i+1}\right),
\end{equation}
and for both \ac{AQC} and \ac{VarQRTE} we use a Hamiltonian variational ansatz.
Gate scheduling follows brickwork pattern: first all odd-even bonds, then all even-odd bonds.
As a baseline, we use a Trotter circuit with twice the number of layers employed by the variational approaches.
\ac{AQC} simulations are carried out with AQC-Tensor~\cite{qiskit-addon-aqc-tensor,Robertson2025_AQC}, while statevector simulations for \ac{VarQRTE} use Qiskit Aer ~\cite{qiskit}.
At each \ac{VarQRTE} step, rather than directly solving the \ac{SLE} in \cref{eq:sle}, we minimize a related quadratic error objective--which improves numerical stability--described in Appendix~\ref{appendix:aqc2mclachlan}, see~\cref{eq:appendix_errorbound}, using \texttt{scipy.optimize.minimize} with \textit{BFGS}. 
The \ac{ODE} is then integrated using a fixed timestep 4th order Runge-Kutta scheme.

\begin{figure}[h!]
\centering
\begin{subfigure}{\columnwidth}
  \includegraphics[width=\textwidth]{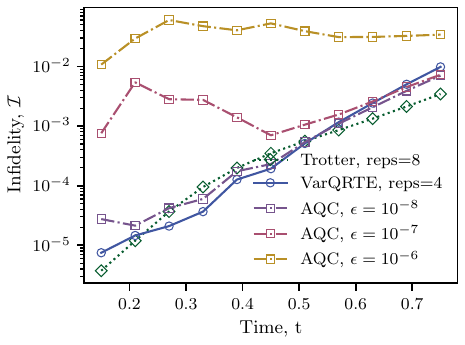}
\end{subfigure}
\begin{subfigure}{\columnwidth}
  \includegraphics[width=\textwidth]{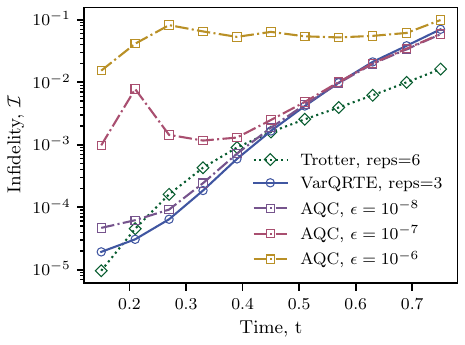}
\end{subfigure}
\caption{Infidelity \(\mathcal{I}\) versus time for AQC (blue), VarQRTE (orange), and Trotter with twice the number of layers (green) on one-dimensional Heisenberg chains. Panel (a) shows results for \(N=35\) with \(L=4\) layers for AQC and VarQRTE and \(L=8\) for Trotter; panel (b) shows results for \(N=50\) with \(L=3\) layers for AQC and VarQRTE and \(L=6\) for Trotter.}
\label{fig:AQC_vs_VarQRTE}
\end{figure}

\cref{fig:AQC_vs_VarQRTE} highlights how the infidelity between the prepared and the target states behaves throughout the evolution for: (a) an $N=35$ qubit system with depth $L=4$ for \ac{AQC} as well as \ac{VarQRTE} and $L=8$ for Trotter, (b) $N=50$ qubits with depth $L=3$ for \ac{AQC} as well as \ac{VarQRTE} and $L=6$ for Trotter. In both cases, \ac{AQC} and \ac{VarQRTE} produce closely matching accuracies across the simulated time window, and both are comparable to the performance of the deeper Trotter circuits. These observations indicate that shallow, fixed-depth variational circuits can approximate the target dynamics with good accuracy to that of substantially deeper product-formula circuits in these regimes.
The close agreement between \ac{AQC} and \ac{VarQRTE} suggests that the attainable accuracy is limited by the expressibility of the chosen ansatz at the prescribed depth. Within the fixed depth, both methods appear to reach similar error floors, consistent with an expressibility-limited regime. This indicates that the bowtie-evaluated \ac{VarQRTE} updates effectively identify parameter trajectories that are near-optimal given the depth constraint.

In the presented examples, \ac{AQC} exhibits a clear accuracy--efficiency trade-off. The \ac{MPS} representation used by \ac{AQC} permits a very stringent truncation threshold (e.g., $10^{-8}$) while maintaining low bond dimension, yielding results that coincide with those of \ac{VarQRTE} and with Trotter at twice the depth. 
However, relaxing the truncation threshold already to $10^{-7}$ leads to a noticeable deterioration of the \ac{AQC} curve relative to \ac{VarQRTE}, as visible in the same figure.
This makes the \ac{AQC} trade-off explicit: efficiency gains from stronger truncation come at a clear accuracy penalty.
In this sense, \ac{AQC} is well suited for simulating deeper circuits, at the price of introducing controlled approximations through tensor truncation.
In contrast, the bowtie/light-cone evaluation used in our \ac{VarQRTE} implementation does not rely on tensor truncation and is therefore best suited for accurate simulations of shallow circuits, with the overall accuracy governed by the ansatz expressibility and by the numerical tolerance of the ODE integration.

In summary, our analysis indicates that \ac{VarQRTE} can achieve similar simulation accuracy as an AQC-compressed Trotter implementation.
While direct Trotter methods are prone to result in deep circuits and the applicability of AQC is limited to systems for which the exact time-evolved state can be calculated or approximated with a sufficiently accurate tensor-network representation, VarQRTE offers us the potential to explore shallow-depth simulation of time-evolved states for which obtaining a trustworthy ground truth would require prohibitive bond dimensions or looser truncation thresholds.
More specifically, for involved system topologies or dimensions, the computational cost of \ac{AQC} can become significant, whereas our approach remains applicable provided the largest bowtie light-cones remain within classical reach. Overall, in regimes where an \ac{MPS} reference is viable, \ac{VarQRTE} matches the accuracy of \ac{AQC} at the same depth and reaches the accuracy envelope of a Trotter circuit with twice the depth, while retaining applicability beyond one-dimensional geometries.
A comparison of the computational resource requirements for \ac{VarQRTE} and \ac{AQC} is given in Section \ref{sec:comp_cost}.

\subsection{State Preparation} \label{sec:SKQD}
Next, we describe an exemplary application of real and imaginary \ac{VarQTE} in the context of sampling-based quantum algorithms for a 2D system. More specifically, \ac{VarQTE} can be employed to construct the circuits representing the time evolution states required in \ac{SKQD} \cite{TheoryQSD22Epperly, yu2025quantumcentricalgorithmsamplebasedkrylov, NoboyukiKSQD25}.

\subsubsection{Sample-based Krylov Quantum Diagonalization}
\ac{SKQD} combines elements from Sample Based Quantum Diagonalization \cite{SQDRobledo_Moreno_2025} and Krylov subspace expansion \cite{Krylov1931}.
This method leverages real-time evolution to generate quantum states to be sampled from and construct a reduced subspace.
Within this subspace, the Hamiltonian is diagonalized to obtain an approximation of concentrated ground states, i.e., states supported on only a polynomial number of bitstrings.
First, we prepare an initial quantum state which needs to have a non-zero overlap with the target ground state.
Secondly, we construct a truncated Krylov subspace by evolving the state according to real time dynamics for different times $\{t_k\}$ and generate the states $\{ \ket{\psi_{t_k}} \}_{k=0}^{M-1}$ which we sample from $L$ times for each $k$. This gives us a set of bitstrings $\left\{s_{kl}|l=0,\ldots,L-1\right\}$, which in turn span the subspace $\mathcal{B}_{ML}=\left\{\ket{s_{kl}}|k=0\ldots M-1;l=0\ldots L-1\right\}$.
Next, we project the target Hamiltonian $H$ onto the subspace $\mathcal{B}_{ML}$. 
Given a subspace whose dimension scales polynomially with the system size, one may diagonalize the projected Hamiltonian exactly using classical computational resources.
For details considering the theoretical guarantees associated with \ac{SKQD}, we refer the interested reader to \cite{yu2025quantumcentricalgorithmsamplebasedkrylov}.
Our \ac{VarQTE} framework can be combined with SKQD as follows: we may employ bowtie \ac{VarQITE} to prepare an initial state with a significant overlap with the ground state
\begin{equation}
    U(\boldsymbol{\theta}^{\tau})\ket{0}^{\otimes n}\approx
    \frac{e^{-H \tau}\ket{\psi_0}}{\sqrt{\bra{\psi_0}-e^{-2H \tau}\ket{\psi_0}}},
\end{equation}
and, subsequently, use bowtie \ac{VarQRTE} to generate the Krylov basis through real-time evolution
\begin{equation}
    U(\boldsymbol{\theta}^{\tau+t_k})\ket{0}^{\otimes n}\approx
     e^{-iH t_k}\frac{e^{-H \tau}\ket{\psi_0}}{\sqrt{\bra{\psi_0}-e^{-2H \tau}\ket{\psi_0}}},
\end{equation}

for $k = 0, \dots, M-1$ by evolving the ansatz parameters. Notably, we may keep the same ansatz $U$ for the VarQITE state preparation and the VarQRTE Krylov basis construction and only vary the parameters $\boldsymbol{\theta}$.
This enables the use of a quantum computer either to directly prepare the respective state and samples thereof or to extend the Krylov basis through additional time simulation steps.

\subsubsection{System Design \& Setup}

Bowtie \ac{VarQTE} provides a particularly natural front-end for \ac{SKQD}. Imaginary VarQTE enables an initialization strategy to prepare a system with a good ground-state overlap and a subsequent real-time VarQTE stage can help to generate a Krylov family $\{\ket{\psi_{t_k}}\}$. 
Moreover, if one wishes to probe real-time evolutions beyond those accessible within the expressibility of ans\"atze compatible with bowtie evaluation, the pipeline extends seamlessly.
A small number of additional Trotter steps can be appended to the end of the variational circuit to reach later times $t_k$, while keeping the early part of the evolution shallow and structured.
In this way, \ac{VarQTE} supplies hardware-friendly Krylov states at low depth, and deeper evolution is introduced only when needed to enrich the Krylov basis.
This preserves the modularity of the overall bowtie-\ac{VarQTE}-enabled \ac{SKQD} workflow.

Next, we introduce the Hamiltonian and the variational ansatz employed in our experimental framework. 
This Hamiltonian has been constructed such that we can tune the sparsity of its ground state. 
Here, \emph{tunable sparsity} refers to the existence of a Hamiltonian parameter that continuously modulates the number of non-negligible amplitudes required for an accurate ground-state representation, thereby allowing a significant fraction to be measured experimentally. 

\begin{figure}[ht]
\centering
\begin{subfigure}{\columnwidth}
    \centering
    \resizebox{0.9\columnwidth}{!}{

\begin{quantikz}[column sep=5pt]
&\gate[label style={black, rotate=90}][0pt][11ex]{\text{Ry}(\theta_0)}
&\gate[label style={black, rotate=90}][0pt][11ex]{\text{Rz}(\theta_2)}
&\gate[2,label style={black, rotate=90}][0pt][11ex]{\text{Rxx}(\theta_4)}
&\gate[2,label style={black, rotate=90}][0pt][11ex]{\text{Ryy}(\theta_5)}
&\gate[2,label style={black, rotate=90}][0pt][11ex]{\text{Rzz}(\theta_6)}
&
\\
&\gate[label style={black, rotate=90}][0pt][11ex]{\text{Ry}(\theta_1)}
&\gate[label style={black, rotate=90}][0pt][11ex]{\text{Rz}(\theta_3)}
&&&&
\end{quantikz}
    }
    \caption{Block of gates to be appied to each pair of qubits.}
    \label{fig:circuitblock}
\end{subfigure}

\vspace{0.2cm}

\begin{subfigure}{\columnwidth}
    \centering
    \includegraphics[width=0.8\columnwidth]{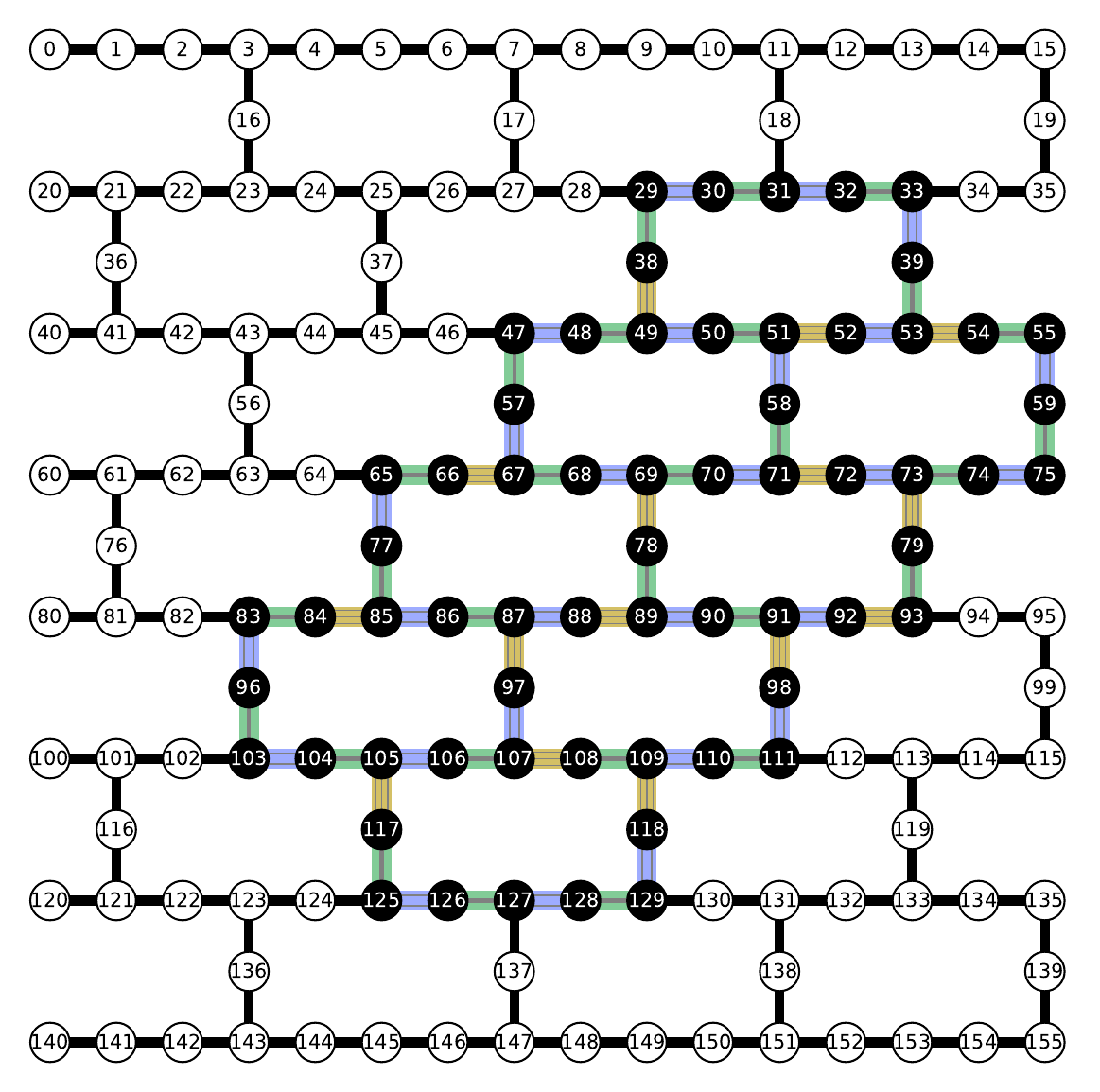}
    \caption{Circuit connectivity. The different color edges correspond to the scheduling of each of the blocks of gates.}
    \label{fig:connectivity}
\end{subfigure}
\caption{Quantum circuit ansatz. The blocks of gates (a) are applied to each colored edge (b) sequentially. The coloring of the edges and the sequence at which we apply the gates have been optimized to maximize expressibility while minimizing the light-cone sizes. If we denote the edges by Blue(\textbf{B}), Orange(\textbf{O}) and Green(\textbf{G}), we found the optimal sequence of gates to be: \textbf{[G,B,O,B,G,B,O]}. Furthermore, a final layer of single qubit rotations is added at the end of the circuit.}
\label{fig:scheduling}
\end{figure}

Our starting point is the antiferromagnetic Heisenberg model on a heavy-hex lattice as shown in~\cref{fig:connectivity}):
\begin{equation}
    H_{\text{Heisenberg}} = - \sum_{\langle i, j \rangle} X_i X_j + Y_i Y_j + Z_i Z_j,
\end{equation}
where $\langle i, j \rangle$ iterates over all nearest-neighbor qubit pairs. This Hamiltonian has a degenerate ground state due to its full rotational symmetry. To obtain a unique, well-defined ground state in the computational basis, we break this symmetry by adding a uniform field:
\begin{equation}
H_{\text{field}} = \sum_{i} Z_i,
\end{equation}
where $i$ runs over all qubits. The combined Hamiltonian $H = H_{\text{Heisenberg}} + c_{\text{field}} H_{\text{field}}$ has a non-degenerate ground state $\ket{0}$ for any $c_{\text{field}} < 0$, which is trivially sparse in the computational basis.
To introduce non-trivial correlations and make the ground state challenging less trivial to describe, we add a disordered perturbation:
\begin{equation}
\label{eq:Hrandom}
    H_{\text{random}} = \sum_{\langle i, j \rangle} \alpha_{i,j} X_i X_j + \beta_{i,j} Y_i Y_j + \gamma_{i,j} Z_i Z_j,
\end{equation}
where the coefficients $\{\alpha_{i,j},\beta_{i,j},\gamma_{i,j}\}_{\langle i,j \rangle}$ are sampled uniformly from $[-1, 1]$. The final Hamiltonian is thus:
\begin{equation}
\label{eq:SKQDHamiltonian}
    H = H_{\text{Heisenberg}} + c_{\text{field}} H_{\text{field}} + c_{\text{random}} H_{\text{random}}.
\end{equation}

By adjusting the strength $c_{\text{random}}$, we can continuously tune the sparsity of the ground state, increasing the number of relevant computational basis states. 
At the same time, the addition of disorder typically leads to more complex many-body states, for example through increased entanglement and the emergence of phenomena such as many-body localization, which are known to pose significant challenges for classical simulation techniques~\cite{tn_disorder, mbl_heisenberg, entanglement_complexity}.

For the design of a variational ansatz which should enable the approximation of the ground state of $H$, we need to consider that our simulation strategy relies on light-cone reduction. Hence, we aim to maximize expressibility while \textit{minimizing the size of the largest light-cone}. We employ a hardware-efficient approach, a generic $\text{SU}(4)$ two-qubit gate as our fundamental building block as is illustrated in~\cref{fig:circuitblock}). We arrange the respective gates on a lattice extracted from the heavy-hex topology of an IBM Heron processor (156 qubits) \cite{qiskit},Green(G), Blue(B) and Orange(O), as shown in \cref{fig:connectivity}.

In order to apply all gates in the circuit within a reasonable overall depth, it is essential to optimize the gate scheduling, i.e., the order in which gates are applied. To this end, we first run a graph-coloring algorithm to partition the set of edges into three mutually non-overlapping groups, as illustrated in \cref{fig:scheduling}. Each group of gates is then applied in sequence following a second-order Suzuki–Trotter decomposition-- that means $[\mathbf{G,B,O,B,G}]$.
Such an ansatz is referred to in literature as a \ac{HVA}: an ansatz resulting of a Trotter circuit for a given Hamiltonian where the time step associated with each Pauli rotation gates is kept as a variational parameters. Such ans\"atze are typically used for real time evolution since its manifold trivially contains the original Trotter circuit and thus inherits all of its mathematical guarantees.
Consider, that in this case, we have added a layer of $\text{R}_\text{y}$ gates to each 2 qubit block. This increases the ansatz expressivity --which is especially important for \ac{VarQITE}--while keeping the size of the lightcones unaffected.
We further optimize the circuit in a greedy fashion by repeating this procedure with different random seeds for the graph-coloring algorithm. In this step, the objective function is the size of the largest light-cone, which we minimize to ensure that the required classical resources remain within feasible limits. The final gate sequence for this circuit is $[\mathbf{G,B,O,B,G,B,O}]$. Including any additional layer would produce light-cones that exceed the statevector simulation capabilities of our available hardware.

From an execution perspective, Qiskit supports \emph{fast parameter updates}, enabling high-throughput execution by repeatedly running the same compiled circuit while updating only its parameter values.
This feature is particularly well suited to parameterized circuits and enables efficient sampling across large parameter sweeps, as discussed in~\cite{fischer_FPU}.
The bowtie-\ac{VarQTE} approach naturally matches this execution model, as the family of Krylov states $\{\ket{\psi_{t_k}}\}$ is represented by a single parameterized ansatz $U(\vtheta)$, with the time dependence encoded solely in the corresponding parameter sets $\{\Vec{\theta}_{t_k}\}$.
As a result, all states required for sampling can be generated using the same compiled circuit, differing only in their parameter values.
Consequently, a bowtie-\ac{VarQTE}-enabled \ac{SKQD} workflow aligns naturally with hardware backends optimized for fast parameter sweeps, enabling efficient acquisition of the full set of required samples.

Finally, we note that the SQD Qiskit add-on \cite{qiskit-addon-sqd} is used for Hamiltonian subspace diagonalization. At the time of writing, this library uses 64-bit integers for bitstring representation for the subspace reduction routine, which restricts the accessible system size to at most 64 qubits. In contrast, the bowtie method could be executed or arbitrarily large system sizes.

\subsubsection{Experiments}
We now present a proof-of-principle realization of the \ac{SKQD} pipeline in which bowtie \ac{VarQITE} is used to prepare an initial state that overlaps with the ground-state. Moreover, bowtie \ac{VarQRTE} is subsequently used to generate a Krylov basis using shallow quantum circuits. Samples collected from these circuits define a compact classical subspace for diagonalization. 
The goal of these experiments is twofold: first, to demonstrate that bowtie \ac{VarQTE} provides a practical and depth-efficient mechanism for generating Krylov basis states that can be fed into \ac{SKQD}; and second, to explore a regime of circuits that may serve as a blueprint for settings in which classical state preparation remains feasible, while obtaining representative samples from the resulting many-qubit states becomes increasingly challenging at scale.

We instantiate the 63 qubit, heavy-hex Hamiltonian of Eq.~\eqref{eq:SKQDHamiltonian} with $c_{\text{field}}=-0.1$ and $c_{\text{random}}=0.15$, where the random couplings in $H_{\text{random}}$ are sampled uniformly as described in \cref{eq:Hrandom}. This parameter choice represents a scenario that yields a ground state which remains sufficiently concentrated in the computational basis to make \ac{SKQD} meaningful, while still exhibiting non-trivial correlations. 
Importantly, at these values the ground-state energy can be computed accurately with tensor-network methods, enabling us to use energy-based benchmarks throughout the pipeline and to verify that the recovered energies trend toward the correct reference value.

\begin{figure}[htbp]
    \centering
    \includegraphics[width=\linewidth]{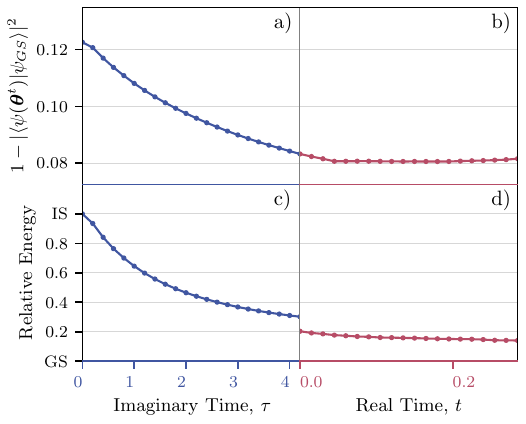}
    \caption{
Energy estimates along the \ac{SKQD} pipeline.
Panel (a) shows the normalized energy $E(\tau)$ of the \ac{VarQITE} trajectory computed with bowtie evaluation, starting from parameters that prepare the first determinant state of the Hamiltonian.
Panel (b) shows the \ac{SKQD} projected-energy estimate obtained from the cumulative support formed by sampling the real-time \ac{VarQRTE} Krylov states $\{ \ket{\psi_{t_k}} \}$: after each additional time point, we construct the reduced Hamiltonian $H_{\mathcal{S}_M}$ and report its ground-state energy.
Energies are rescaled such that $E_{\mathrm{GS}}=0$ and $E_{\mathrm{IS}}=1$.
Real-time evolution uses time step $\Delta t=\pi/\Delta E=0.015$ and each Krylov state is sampled with $N_{\mathrm{shots}}=10^4$ shots.
}
    \label{fig:SQDPipeline}
\end{figure}

~\cref{fig:SQDPipeline} summarizes the end-to-end \ac{VarQTE}+\ac{SKQD} workflow for such a Hamiltonian and using the ansatz in \cref{fig:scheduling}.
The 4 different panels show the evolution of the infidelity between the variational state and our reference computation of the ground state(panels (a) and (b) on top), and our energy estimation(panels (c) and (d) at the bottom).
The energy scale has been normalized such that the energy of the initial state (indicated as IS)---also the leading determinant of the Hamiltonian--- is 1 and the reference ground state energy (indicated as GS) is 0.
The plot is further divided between the imaginary time evolution ((a) and (c)--blue) and real time evolution ((b) and (d)--red).
For imaginary time evolution, the energy estimate is given by the exact energy of the variational state, which is computed via the bowtie method; for real time evolution, the energy estimate is given by sampling $10^5$ times from all the Krylov states up to time $t$, projecting the Hamiltonian into the subspace of sampled bitstrings, and diagonalizing it.

During the imaginary time evolution, the infidelity with the ground state decreases (a) and the estimated energy decreases significantly (c). 
In fact, for the last state in the imaginary time trajectory we have computed an infidelity of $1.5\times10^{-2}$ with the initial state. 
Sampling from this state produces a non-trivial bitstring set that can be used to define a classical subspace. 
We observe a discontinuity between the final point of the imaginary-time \ac{VarQITE} trajectory and this initial \ac{SKQD} estimate, even though both originate from the same underlying quantum state. 
This difference arises because the two energies are computed in fundamentally different ways. 
The \ac{VarQITE} energy is given by the variational expectation value 
$E_{\text{VarQITE}} = \langle \psi(\theta) | H | \psi(\theta) \rangle$, 
which evaluates the energy of a single state. 
In contrast, the \ac{SKQD} energy is obtained by sampling bitstrings from this state, constructing a subspace spanned by these configurations, and diagonalizing the Hamiltonian within that subspace.
As a consequence, even in the absence of Krylov expansion, \ac{SKQD} effectively performs a local optimization within the support of the state. 
This allows the sampled configurations to be recombined into a state with lower energy than the original variational wavefunction, leading to the observed gap. 
In the present Hamiltonian instance, the projected energy is therefore already lower than the variational energy of the same \ac{VarQITE} state.
We emphasize, however, that this behavior is instance dependent: depending on how the probability weight is distributed over computational basis states and how well the sampled configurations capture low-energy components, the projected estimate from a single state may coincide with or even be worse than the corresponding variational energy.

During the real time evolution (d), we can see that as we increased the number of Krylov states the recovered energy improves further, consistent with the expected monotonicity of enlarging the classical subspace: adding additional sampled bitstrings and/or additional Krylov states cannot worsen the best achievable energy within the projected subspace.
We sampled up to 20 Krylov states, with $10^4$ measurement shots collected for each state.
Note that we have chosen the timestep for the Krylov dynamics in agreement with existing theory \cite{skqdibm} as
\begin{equation}
    \Delta t= \frac{\pi}{\Delta E},
\end{equation}
where $\Delta E=E_1-E_0$ corresponds to the difference between the smallest eigenvalues of $H$. The value of $\Delta E$ has been roughly approximated using MPS simulation.
A practical limitation of the present experiment is that the ground state is very concentrated in the computational basis: a single dominant bitstring carries a large fraction of the total probability mass, and this dominance persists across a substantial portion of the generated Krylov states. As a consequence, repeated sampling spends a large fraction of the shot budget re-observing the same configuration, and only a comparatively small fraction of measurements contributes genuinely new bitstrings to $\mathcal{S}_M$. 
For the present Hamiltonian, the bitstring distribution is strongly peaked, so the number of \emph{new} configurations discovered per shot is small and the cumulative sampled support grows slowly with additional shots. In turn, the projected subspace used by \ac{SKQD} expands only gradually, motivating the search for instances whose concentrated ground states distribute weight more broadly across their support to improve sample efficiency.

\subsection{Computational Cost} \label{sec:comp_cost}

Next, we evaluate the cost of bowtie \ac{VarQTE} with respect to the gradient vector $b(\vtheta)$ and the \ac{QGT} $g(\vtheta)$.
The computational resource requirements of the method are governed by: (i) the simulation of the bowtie states $\{\tb{i}\}_{i=0}^{N_p-1}$, $\{\tc{i}\}_{i=0}^{N_h-1}$, and (ii) the computation of the overlaps given in \cref{eq:overlaps1,eq:overlaps2,eq:overlaps3,eq:overlaps4}.
The number of bowtie states $\{\tb{i}\}_{i=0}^{N_p-1}$ and $\{\tc{i}\}_{i=0}^{N_h-1}$ scales as $\mathcal{O}\left(N_p\right)$ and $\mathcal{O}\left( N_h\right)$, respectively.
The number of overlaps scales with $\mathcal{O}\left(N_p (N_p+N_h)\right)$.
Nevertheless, we observe in our experiments that the wall-clock time is dominated by the computation of the bowtie states rather than the overlaps.
The moderate resource costs required to calculate the overlaps of the form $\braket{s_A}{s_B}$ are often due to the limited numbers of qubits which appear within the light-cones of $\ket{s_A}$ and $\ket{s_B}$. Firstly, many overlaps are trivially zero because the set of qubits within the light-cones of the respective states are disjoint. In our examples, only about $20\%$ of all matrix entries in the QGT and the gradient require non-trivial overlap computation.
Secondly, the overlap computation between two bowtie states $\ket{s_A}$ and $\ket{s_B}$ supported on different sets of qubits can be optimized.
More specifically, the overlap computation does not need to be executed for the full vector space. Instead one may identify the intersection between the set of qubits $n_A$ and $n_B$ in the light-cones of $\ket{s_A}$ and $\ket{s_B}$ with overlapping action and limit the evaluation to a sparse vector multiplication in the reduced subspace. Depending on the system, this can lead to significant reductions in the computational cost. Additional details are given in Appendix~\ref{app:overlap_qubit_intersect}.
\begin{figure}[h!]
    \centering
    \includegraphics[width=\linewidth]{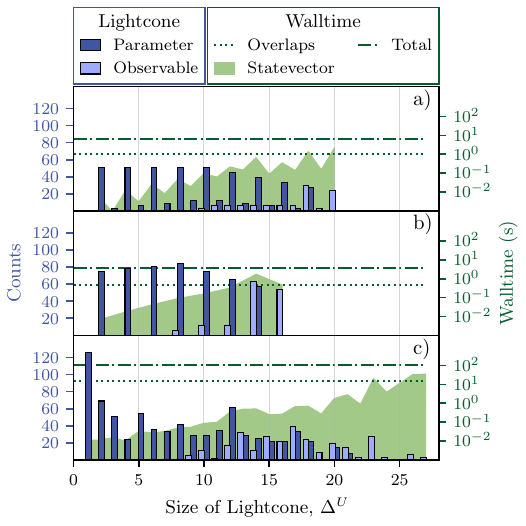}
    \caption{Resources needed for the computation of variational quantities using the bowtie method. We construct all bowties for a VarQTE routine with a given quantum circuit and observable. On the left vertical axis we plot the distribution of the lightcone sizes for each bowtie for both $B_{\beta}$(in blue) and $B_{\gamma}$(in yellow). With the right vertical axis we plot the runtime to simulate all bowtie circuits of a given size using statevector simulation. (a) and (b) correspond to 1D lattices, while (c) corresponds to a heavy-hex topology.}
    \label{fig:BowtieCost}
\end{figure}
An overview of the computational resources required in our experiments is given with respect to the light-cone sizes and the runtimes in~\cref{fig:BowtieCost}. 
We repeat the results for 3 different systems. a) and b) correspond to the quantum circuits used for AQC in Sec.~\ref{sec:comparaison} for N=35 and N=50 respectively, and c) corresponds to the circuit introduced in \cref{fig:scheduling}.
To analyze the cost of computing the QGT and the gradient individually, we extract both $\{B_{\beta}^{i}\}_{i=0}^{N_p-1}$ and $\{B_{\gamma}^{j}\}_{j=0}^{N_h-1}$ and classify them according to their lightcone size~$\Delta^U$.
The $x$-axis reports the lightcone sizes~$\Delta^U$ and the left $y$-axis indicates the number of occurrences of the different light-cones sizes. 
The two main observations from the histogram plot are that (i) the $\Delta_{\beta}$ distribution is concentrated around smaller sizes, whereas the $\Delta_{\gamma}$ are consistently larger--with the largest lightcones always arising uniquely from $B_{\gamma}$--and (ii) large bowties occur far less frequently than small ones, especially once we depart from trivial 1D topologies.
The right $y$-axis indicates the runtime associated with computing the QGT.
First, we observe that overlap computation is roughly an order of magnitude cheaper than the total runtime--which is illustrated as a dotted line.
We plot the time required to simulate all bowtie states of a given lightcone size as a shaded green area.
The majority of the total simulation time is spent on a few bowtie states corresponding to larger light-cones.
This behavior is expected, as the cost of statevector simulation grows exponentially with the system size.
Notably, overlaps which include bowtie states whose light-cone is too large to be simulated efficiently classically, could also be directly evaluated on a quantum computer.
Furthermore, we observe larger light-cones for the $\{\tc{j}\}$ states compared to the bowtie states $\{\tb{i}\}$. 
Since the former are only needed to evaluate the gradient $\vec{b}(\vtheta)$ but not the QGT $g(\vtheta)$, the runtimes of our experiments for evaluating $\vec{b}(\vtheta)$ end up significantly higher than for $g(\vtheta)$, although the latter have more entries.
All wall-clock timings in this section were obtained on an 11th Gen Intel\textsuperscript{\textregistered} Core\texttrademark{} i7-11850H @ 2.50\,GHz machine with 64\,GB RAM, using 10 parallel threads. We employ the \textit{Qiskit Aer} simulator in \textit{statevector} mode with a \textit{CPU} backend; parallelization of circuit simulations is handled by the simulator’s default settings. To extract size-resolved runtimes we batch bowties by lightcone size, whereas in practice one would typically submit all bowties concurrently, which improves hardware utilization and reduces overhead. The reported total-runtime curves correspond to executing the complete set of bowties without this per-size batching.
\\

Furthermore, we compare the computational cost of running VarQTE and AQC--see Section \ref{sec:expressivity}.
Our simulations were mainly constrained by two factors.
First, the largest exact statevector simulation that we could execute reliably on our cluster involved $27$ qubits. This imposed a hard limit on the circuit sizes and depths that could be explored, although this boundary is ultimately hardware dependent. At this point it is worth highlighting a recent milestone, where a full exact simulation of a universal $50$-qubit system has been achieved on the JUPITER supercomputer; see~\cite{deraedt2025universalquantumsimulation50}.
Second, our AQC implementation relies on the software library \cite{AQC_tensor}, whose tensor-network backends are primarily optimized for one-dimensional circuit topologies. While more sophisticated tensor-network methods could in principle extend the applicability of AQC to non-trivial higher-dimensional layouts, incorporating such techniques lies beyond the scope of this work.

We restrict ourselves to a theoretical resource analysis of circuit-simulation backends and use it to compare the dominant simulation costs entering VarQTE and AQC. Throughout this discussion, we estimate the number of \ac{FLOPs} required for a single VarQTE time step and for a single \ac{AQC} optimization step, assuming a fixed ansatz and a fixed observable. Furthermore, we consider only the cost associated with simulating the circuits in both methods: bowtie circuits for bowtie VarQTE and the full circuit for \ac{AQC}. This single-step analysis does not capture potential differences in the total number of iterations required in practice and neglects additional computational overheads; nevertheless, it provides a useful proxy for comparing the relative computational costs of the two approaches. Notably, by neglecting the cost of the initial target-state preparation in \ac{AQC}, our comparison is favorable to that method.

Let $G$ denote the number of gates in the compiled circuit, under the assumption that all gates are two-qubit gates; let $n$ denote the number of qubits, and let $\chi$ denote the maximum bond dimension used by a tensor-network backend.
To obtain concrete numerical estimates, rather than scaling laws up to unknown constants, we use reduced \ac{FLOPs} models that retain the dominant contraction terms and set implementation-dependent prefactors to one. The resulting backend-dependent formulas are summarized in Table~\ref{tab:backend_costs}.

\begin{table}[ht]
    \centering
    \caption{Surrogate \ac{FLOPs} models for a compiled circuit consisting of $G$ two-qubit gates acting on $n$ qubits in a topology with coordination number $z$. Here $\chi$ denotes the maximum bond dimension. These expressions retain the dominant local-update terms and are intended as concrete cost estimates for relative comparisons.}
    \label{tab:backend_costs}
    \small
\begin{tabular}{|c|c|}
\hline
        \textbf{Backend} & \textbf{FLOP model} \\
        Statevector
        & $F_{\mathrm{SV}}(n,G)=16\,G\,2^n$ \cite{qelvin_statevector,qiskit_statevector} \\[0.4em]

        MPS
        & $F_{\mathrm{MPS}}(G,\chi)=16\,G\,\chi^3$ \cite{vidal2003slightlyentangled,qiskit_mps,tensorsnet_mps_tebd} \\[0.4em]

        PEPS
        & $F_{\mathrm{PEPS}}(G,\chi)=G\left(4\chi^{z+1}+64\chi^3\right)$ \cite{lee2025peps_rqc} \\
        \hline
    \end{tabular}
\end{table}

The statevector expression follows from the standard gate-application picture in which the full wavefunction contains $2^n$ amplitudes and each dense two-qubit gate has $16$ distinct amplitudes. The MPS expression follows from the standard TEBD local-update cost $d^4\chi^3$--where the dimension of a qubit is $d=2$.
The PEPS expression considers a simple-update cost: the bond-projection step scales as $\chi^{z+1}d^2$ and the local SVD step as $\chi^3d^6$ where $z$ is the coordination number--$z=3$ for a heavy-hex lattice and $z=4$ for a square lattice.
These formulas should be interpreted as concrete surrogate FLOP counts rather than as exact backend-specific operation counts. Their purpose is to provide a uniform basis for comparing circuit-simulation cost across backends within the simplified single-step resource model adopted here.

\begin{table*}[ht]
\begin{tabular}{|c|ccc|cc|}
\hline
\multirow{2}{*}{} & \multicolumn{3}{c|}{VarQTE} & \multicolumn{2}{c|}{AQC}                             \\ \cline{2-6} 
                  & \multicolumn{1}{c|}{SV}              & \multicolumn{1}{c|}{TN(64)}        & \multicolumn{1}{c|}{TN(256)}  & \multicolumn{1}{c|}{TN(64)}        & TN(256)       \\ \hline
Square Lattice Heisenberg (1)& \multicolumn{1}{c|}{$1.4\times 10^{12}$}                  & \multicolumn{1}{c|}{$2.8\times 10^{12}$ }               & \multicolumn{1}{c|}{$6.1\times 10^{14}$}                 & \multicolumn{1}{c|}{$4.5\times 10^{10}$}               &  $9.9\times 10^{12}$\\ \hline
Square Lattice Ising (2)& \multicolumn{1}{c|}{$1.9 \times 10^{10}$}                  & \multicolumn{1}{c|}{$1.3\times 10^{12}$}               & \multicolumn{1}{c|}{$2.9\times 10^{14}$}                   & \multicolumn{1}{c|}{$4.7\times 10^{10}$}               & $1.0\times 10^{13}$               \\ \hline
Heavy Hex Heisenberg (2) & \multicolumn{1}{c|}{$5.5\times 10^{12}$}                  & \multicolumn{1}{c|}{$8.3\times 10^{11}$}               & \multicolumn{1}{c|}{$5.3\times 10^{13}$}                   & \multicolumn{1}{c|}{$1.4\times 10^{ 10}$}               &$9.0\times  10^{11}$                \\ \hline
1D Heisenberg (3)& \multicolumn{1}{c|}{$8.9 \times 10^{9}$} & \multicolumn{1}{c|}{$4.8\times10^{11}$} & \multicolumn{1}{c|}{$3.07\times10^{13}$} &     \multicolumn{1}{c|}{$8.67\times 10^9$} & $5.54\times10^{11}$ \\ \hline
\end{tabular}
\caption{Estimated \ac{FLOP} counts for circuit simulation using the Bowtie (VarQTE) method and \ac{AQC}, evaluated for four representative systems with different lattice geometries, Hamiltonians, and Hamiltonian variational ansätze (HVA). For VarQTE, the reported cost corresponds to the sum of the simulation costs of all bowtie circuits. For \ac{AQC}, the cost corresponds to simulating the full variational circuit. Estimates are reported for a statevector (SV) simulator and for PEPS tensor-network simulators with bond dimensions $\chi=64$ and $\chi=256$. All values reflect a single VarQTE time step or a single \ac{AQC} optimization step and account only for circuit-simulation costs.}
\label{tab:flop}
\end{table*}

In \cref{tab:flop}, we report estimated \ac{FLOP} counts obtained using the Bowtie method and \ac{AQC} for a range of Hamiltonians and lattice geometries. Under the VarQTE column, we estimate the computational resources by summing the \ac{FLOP} costs required to simulate each of the individual bowtie circuits. We consider three circuit-simulation backends: a statevector (SV) simulator, as well as two PEPS tensor-network (TN) simulators with bond dimensions $\chi=64$ and $\chi=256$.
Under the \ac{AQC} column, we estimate the cost of simulating the full variational circuit using PEPS tensor networks with the same bond dimensions. In all cases, the reported values correspond exclusively to the cost of circuit simulation, consistently with the resource-analysis framework introduced above.
We repeat this cost estimation for four representative systems. For each system, we specify a lattice topology, a local Hamiltonian, and a variational ansatz chosen to be a Hamiltonian variational ansatz (\ac{HVA}), following the same construction as in Section \ref{sec:SKQD}. The Hamiltonians defined on the lattice take one of the following forms:
\begin{equation}
H_{\mathrm{H}} = \sum_{\langle i,j \rangle} \left( X_i X_j + Y_i Y_j + Z_i Z_j \right),
\end{equation}
or
\begin{equation}
H_{\mathrm{I}} = \sum_{\langle i,j \rangle} Z_i Z_j + \sum_{i=0}^{N-1} X_i,
\end{equation}
where $\langle i,j \rangle$ runs over all edges of the underlying interaction graph.
The four systems considered are as follows:
\begin{itemize}
    \item A $10\times10$ square lattice with a Heisenberg Hamiltonian, using an HVA with a single variational layer.
    \item A $10\times10$ square lattice with an Ising Hamiltonian in a transverse field, using an HVA with two variational layers.
    \item A Heavy-Hex lattice with 114 qubits and a Heisenberg Hamiltonian, using an HVA with two variational layers.
    \item A one-dimensional lattice with 50 qubits and a Heisenberg Hamiltonian,using an HVA with 3 layers, for which both the Hamiltonian and variational circuit coincide with those used in Section \ref{sec:comparaison}.
\end{itemize}

The results in \cref{tab:flop} highlight two main trends. First, for the classes of circuits considered here, simulating the collection of bowtie circuits using tensor-network backends does not reduce the computational cost relative to simulating the corresponding full variational circuit once. In particular, when comparable tensor-network representations are employed, the aggregate cost of contracting all bowtie circuits exceeds that of a single contraction of the full circuit. This observation reflects the overhead associated with repeating the tensor-network simulation of the same regions of the circuit multiple times.
Second, for the system sizes studied in this work—where the largest lightcones induced by the bowtie decomposition remain moderately sized—the Bowtie method yields a consistent computational advantage. In this regime, decomposing the dynamics into localized circuits leads to a reduction in the effective problem size that outweighs the cost of repeated simulations, resulting in lower overall resource requirements.
We note that it is currently unclear whether the existence of a favorable bowtie decomposition implies the existence of an efficient tensor-network representation, and clarifying this relationship remains an open question.

Furthermore, while our comparison is based exclusively on \ac{FLOP} counts, it is important to note that this metric does not fully capture practical performance differences between simulation backends. In practice, statevector simulations often benefit from more regular memory access patterns, better cache utilization, and highly optimized linear-algebra kernels, which can translate into superior wall-clock performance on classical hardware. Consequently, although tensor-network and statevector simulators may exhibit similar nominal \ac{FLOP} counts in certain regimes, their practical execution times can differ substantially.

\section{Discussion and Outlook}
\label{sec:discussion}

Real and imaginary time evolution are native approaches for preparing physically structured states--with the latter being particularly suited for warm-starting ground state search algorithms. This is because imaginary time evolution comes with interesting guarantees (in the limit) that provide a theoretical foundation for preparing ground states.
Now, this work presents a resource-efficient framework for the preparation of quantum states via real or imaginary time evolution enabled by causal-lightcone based classical simulation--\textit{bowtie \ac{VarQTE}}. 
We introduce a framework for \ac{VarQTE} that employs classical simulation methods and, thereby, addresses a central bottleneck: the reliable and efficient evaluation of the quantities entering McLachlan's variational equations. By exploiting the structure of the variational ansatz and identifying causal light-cones, we decompose global evolution steps into smaller independent subproblems that are classically tractable. This enables exact classical evaluation of the \ac{QGT} and gradient terms for relevant system sizes, while reserving quantum resources for tasks where they are truly required such as sampling.
It is important to note that the states prepared with bowtie \ac{VarQTE} can always be directly mapped onto a quantum computer and, thus, processed further with a quantum algorithm of choice.
Furthermore, the framework is inherently modular and can be seamlessly combined with other classical or quantum techniques. For instance, tensor-network contractions or quantum hardware can also be employed selectively to evaluate specific \ac{QGT} or gradient elements that are intractable for causal-lightcone simulations.

\ac{VarQTE} provides a promising state preparation primitive that is particularly well-suited if one of the following settings is given: 
\begin{itemize}
    \item An approximate imaginary-time evolution can drive the system towards a state that has a significant overlap with the target state and, hence, provides a high-quality warm start, e.g., for ground state search algorithms such as QAOA or QPE.
    \item A short-time variational evolution helps to generate an initial state which may be combined with long-time, coherent simulation executed, e.g., with a Trotter method on a quantum computer. This workflow is compatible with Hamiltonian simulation and thermal state preparation. A key challenge in this direction is the model-selection trade-off: one must choose Hamiltonians with sufficiently local terms to support efficient \ac{VarQTE}, while still exhibiting concentrated ground states, which are required for \ac{SKQD} to perform well.

\end{itemize} 

A central objective of future research on bowtie VarQTE should be the preparation of more complex states. Achieving this goal might require extending the method beyond the limits of light-cone state-vector simulation and integrating tensor-network techniques or Pauli propagation into the workflow.
A particularly compelling path is a hybrid strategy: use exact state-vector simulation for small bowties, and selectively employ tensor networks and Pauli propagation for the largest bowties---especially $B_{\gamma}$, which, in our experiments, are typically the largest and contribute to gradient terms that are the least sensitive to noise in the \ac{SLE} solve.

Finally, bowtie VarQTE presents a primitive whose properties complement existing state preparation protocols and may, therefore, provide a practical route toward resource-efficient initialization of physically structured quantum states relevant for sampling and simulation tasks.

\paragraph*{Acknowledgments}
We thank Julien Gacon and Ben Jaderberg for insightful discussions.
This work was supported by the Swiss State Secretariat for Education, Research and Innovation (SERI) under SBFI No. UeM019-10.2.

\paragraph*{Code availability}
To facilitate transparency and reproducibility, a reference implementation of the proposed method, together with illustrative examples, has been released as open-source software. The codebase is publicly accessible via the repository described in \cite{zenododrudis}.

\newpage

\bibliographystyle{apsrev4-1}
\bibliography{references}

\newpage

\onecolumngrid
\appendix

\section{Alternative Strategies for Estimating the QGT}
\label{app:alternative_qgt_strategies}

The efficient evaluation of the \ac{QGT} is one of the main computational bottlenecks in the application of \ac{VarQTE}. 
For small problem instances, where full state-vector simulation remains feasible, the \textit{reverse-gradient} technique introduced in \cite{reverseqgt} provides an exact evaluation method. 
This approach exploits the internal structure of the \ac{QGT} to reuse intermediate states, thereby reducing the computational cost of the required vector multiplications. 
To the best of our knowledge, it remains the fastest exact method for simulating \ac{VarQTE} on circuits of arbitrary depth. 
However, its applicability is limited to system sizes for which state-vector simulation is still possible.

For larger systems, exact \ac{QGT} evaluation is generally too expensive, and the literature has therefore considered approximate alternatives. 
A prominent class of such approaches replaces the explicit construction of all \ac{QGT} entries by stochastic estimators, thereby avoiding the direct quadratic scaling in the number of variational parameters. 
In this appendix, we first summarize representative stochastic approximation methods from the literature. 
We then discuss two classical strategies that we considered for evaluating the fidelity terms appearing in one such estimator, namely \ac{SPSA}. 
Both strategies exploit the fact that the two variational states whose overlap is required differ only by a small perturbation of the circuit parameters. 
Finally, we show that, even under the optimistic assumption that these fidelity terms can be evaluated exactly, the number of stochastic samples required to obtain a \ac{QGT} estimate of sufficient accuracy is prohibitively large for the \ac{VarQTE} applications considered in this work.

\subsection{Related Work: Stochastic Approximation Methods}
\label{app:related_work_stochastic_qgt}

Several strategies have been proposed to reduce the quantum resources required to calculate the \ac{QGT}.
Among these, stochastic approximation methods are particularly promising because they avoid the quadratic scaling in the number of parameters. 
We focus on three representative approaches.

The method developed in \cite{shadows} reformulates the problem in the Heisenberg picture. 
Instead of simulating the imaginary time evolution of a quantum state, one projects its density matrix onto a set of operators $\mathcal{S}$ and formulates the dynamics in terms of their expectation values.
Under the assumption that the variational state is well described by a limited number of local operators, a variational principle derived from Ehrenfest's theorem yields a formally analogous object to the \ac{QGT} in \cref{eq:qgt},
\begin{equation}
  g_{i,j}(\vtheta,S)
  =
  \sum_{O\in \mathcal{S}} 
  \partial_i \langle O \rangle_{\vtheta}^* \,
  \partial_j \langle O \rangle_{\vtheta}.
\end{equation}
This tensor can be constructed by measuring expectation-value gradients of a chosen set of observables, leading to a cost that scales only linearly with the number of parameters.
The method is particularly well suited for execution on quantum computers, where multiple observables can be measured efficiently using classical shadows.

A second approach uses \ac{SPSA} \cite{spsa} to obtain an unbiased estimator for the \ac{QGT}. 
By sampling parameter perturbations $\spsapert$ from a suitable distribution $\mathcal{D}$, the \ac{QGT} can be approximated as
\begin{equation}
  \Hat{g}(\vtheta)
  =
  \sum_{\spsapert_1,\spsapert_2 \sim \mathcal{D}}
  -\frac{\delta F(\vtheta,\spsapert_1,\spsapert_2)}{4 \epsilon^2}
  \frac{\spsapert_1\spsapert^T_2+\spsapert_1^T\spsapert_2}{2},
\end{equation}
where $\epsilon$ is a small perturbation parameter and
\begin{align}
  \label{eq:approx_fidelity}
  \begin{split}
    \delta F(\vtheta,\spsapert_1,\spsapert_2)
    &=
    \abs{\langle \psi(\vtheta) | \psi(\vtheta+\epsilon \spsapert_1+\epsilon \spsapert_2) \rangle }^2 \\
    &\quad
    -
    \abs{\langle \psi(\vtheta) | \psi(\vtheta+\epsilon \spsapert_1) \rangle }^2 \\
    &\quad
    -
    \abs{\langle \psi(\vtheta) | \psi(\vtheta-\epsilon \spsapert_1+\epsilon \spsapert_2) \rangle }^2 \\
    &\quad
    +
    \abs{\langle \psi(\vtheta) | \psi(\vtheta-\epsilon \spsapert_1) \rangle }^2 .
  \end{split}
\end{align}
A principal advantage of this method lies in its allocation of a fixed shot budget. 
Given a finite number of total shots, one can distribute them across the required circuit evaluations. 
The estimator is statistically efficient when using just one shot per circuit, thereby maximizing the number of distinct parameter perturbations sampled. 
Moreover, the estimate is incrementally refinable: after an initial budget is spent, its precision can be systematically improved by allocating additional shots.

In a similar spirit, the method introduced in \cite{steinqgt} exploits a connection between stochastic \ac{QGT} estimation and Stein's identity. 
While \ac{SPSA} samples perturbations from the vertices of a hypercube in parameter space, this approach instead uses perturbations drawn from a normal distribution.

\subsection{Classical Evaluation of the SPSA Fidelity Terms}
\label{appendix:AlternativeQGTComputing}

The \ac{SPSA} estimator reduces the problem of approximating the \ac{QGT} to repeated evaluations of the fidelity between a variational state at parameters $\vtheta$ and nearby states obtained by perturbing those parameters. 
This observation suggests a possible hybrid strategy: rather than evaluating the fidelities on quantum hardware, one may attempt to compute them with approximate classical simulation methods tailored to the special structure of the overlap. 
In the present setting, the two circuits entering each overlap have identical architecture and differ only by small parameter shifts. 
This proximity can be exploited to reduce the approximation error while keeping the computational cost lower than that of a generic overlap computation.

We considered two such classical strategies. 
The first is based on Pauli propagation and uses the fact that small changes in the rotation angles can be incorporated directly into the propagation rules. 
The second is based on a middle-out tensor-network contraction strategy, which aims to exploit the similarity between the two circuits when computing their overlap. 
Although both methods are conceptually compatible with the \ac{SPSA} estimator, we ultimately did not use them in the main numerical workflow. 
The reason is that the dominant obstruction is not only the cost of estimating an individual fidelity. 
As shown in Section ~\ref{section:ComputingTheQGTAccurately}, even an idealized version of \ac{SPSA} with exact fidelity evaluations requires an impractically large number of random directions to reach the accuracy needed for stable \ac{VarQTE} dynamics.

\subsubsection{Pauli Propagation with different angles}
\label{sec:unevenPP}

We first consider the propagation of a Pauli operator $Q$ under a Pauli rotation
\begin{equation}
  R_P(\theta)=e^{-i\frac{\theta}{2}P},
\end{equation}
where the rotation angles on the two sides differ by a small shift. 
The relevant object is
\begin{align}
R_P(\theta+\Delta \theta) Q R_P^\dag(\theta-\Delta \theta)
&=
\cos\left(\frac{\theta+\Delta \theta}{2}\right)
\cos\left(\frac{\theta-\Delta \theta}{2}\right) Q \nonumber\\
&\quad+
\sin\left(\frac{\theta+\Delta \theta}{2}\right)
\sin\left(\frac{\theta-\Delta \theta}{2}\right) P Q P \nonumber\\
&\quad+
\cos\left(\frac{\theta+\Delta \theta}{2}\right)
\sin\left(\frac{\theta-\Delta \theta}{2}\right) i Q P \nonumber\\
&\quad-
\sin\left(\frac{\theta+\Delta \theta}{2}\right)
\cos\left(\frac{\theta-\Delta \theta}{2}\right) i P Q .
\end{align}
The expression simplifies depending on whether $P$ and $Q$ commute or anticommute.

If $[P,Q]=0$, then
\begin{align}
& R_P(\theta+\Delta \theta) Q R_P^\dag(\theta-\Delta \theta) \nonumber\\
&=
\left[
\cos\left(\frac{\theta+\Delta \theta}{2}\right)
\cos\left(\frac{\theta-\Delta \theta}{2}\right)
+
\sin\left(\frac{\theta+\Delta \theta}{2}\right)
\sin\left(\frac{\theta-\Delta \theta}{2}\right)
\right] Q \nonumber\\
&\quad+
\left[
\cos\left(\frac{\theta+\Delta \theta}{2}\right)
\sin\left(\frac{\theta-\Delta \theta}{2}\right)
-
\sin\left(\frac{\theta+\Delta \theta}{2}\right)
\cos\left(\frac{\theta-\Delta \theta}{2}\right)
\right] i P Q \nonumber\\
&=
\cos(\Delta \theta) Q - \sin(\Delta \theta) i P Q .
\end{align}

If $\{P,Q\}=0$, then
\begin{align}
& R_P(\theta+\Delta \theta) Q R_P^\dag(\theta-\Delta \theta) \nonumber\\
&=
\left[
\cos\left(\frac{\theta+\Delta \theta}{2}\right)
\cos\left(\frac{\theta-\Delta \theta}{2}\right)
-
\sin\left(\frac{\theta+\Delta \theta}{2}\right)
\sin\left(\frac{\theta-\Delta \theta}{2}\right)
\right] Q \nonumber\\
&\quad+
\left[
\cos\left(\frac{\theta+\Delta \theta}{2}\right)
\sin\left(\frac{\theta-\Delta \theta}{2}\right)
+
\sin\left(\frac{\theta+\Delta \theta}{2}\right)
\cos\left(\frac{\theta-\Delta \theta}{2}\right)
\right] i P Q \nonumber\\
&=
\cos(\theta) Q + \sin(\theta) i P Q .
\end{align}

The different-angle propagation rule provides a way to express the overlap between two parameterized circuits that share the same structure but differ by small parameter shifts. 
This is precisely the setting encountered in the fidelity terms of the \ac{SPSA} estimator. 
In this context, Pauli propagation is appealing because the difference between the two circuits is localized in parameter space and enters through the small shifts $\Delta\theta$. 

To the best of our knowledge, this different-angle formulation has not previously been implemented within a Pauli-propagation framework. 
Consequently, there is currently no established baseline for its practical performance or for the regimes in which it may provide an advantage. 
Depending on the observed growth of the propagated Pauli representation, this approach could nevertheless provide an interesting route to obtaining rough estimates of the \ac{QGT}, especially in settings where approximate fidelity evaluations are sufficient. 
If the resulting Pauli representation remains sufficiently compact, this method could be particularly efficient for circuits with few but highly entangling gates.

\subsubsection{Middle-out MPO construction}

We consider the problem of evaluating overlaps between two nearby parameterizations of the same quantum circuit. This can be expressed as the expectation value of an operator of the form $U^\dagger(\vtheta) U(\vtheta+\delta\vtheta)$, which admits a natural representation as a two-layer tensor network. The efficient contraction of such structures using MPO techniques has been explored, for instance, in the context of error mitigation \cite{filippov2024scalabilityquantumerrormitigation}, and we adapt this idea here to variational quantum dynamics. 
Given a parameterized quantum circuit
\begin{equation}
U(\vtheta) = \prod_{j=0}^{N_p-1} U_j(\theta_j),
\end{equation}
and the overlap between two nearby parameterizations
\begin{equation}
\braket{\psi(\vtheta)}{\psi(\vtheta+\delta\vtheta)}
=
\bra{0} U^\dagger(\vtheta)\, U(\vtheta+\delta\vtheta) \ket{0}.
\end{equation}
Using the circuit decomposition, this can be written as
\begin{equation}
U^\dagger(\vtheta)\, U(\vtheta+\delta\vtheta)
=
\left(\prod_{j=0}^{N_p-1} U_j^\dagger(\theta_j)\right)
\left(\prod_{j=N_p-1}^{0} U_j(\theta_j+\delta\theta_j)\right).
\end{equation}

We construct this operator iteratively as a matrix product operator (MPO). Starting from the identity
\begin{equation}
\widetilde{U}_{N_p} = \mathbb{I},
\end{equation}
we define a sequence of operators
\begin{equation}
\widetilde{U}_j
=
U_j^\dagger(\theta_j)\, \widetilde{U}_{j+1}\, U_j(\theta_j+\delta\theta_j),
\quad j = N_p-1, \dots, 0,
\end{equation}
so that $\widetilde{U}_0 = U^\dagger(\vtheta)\, U(\vtheta+\delta\vtheta)$.

At each step, a pair of gates is inserted on both sides of the current operator. The MPO is therefore grown around an initial identity by iterating over the circuit from the last gate to the first. For small parameter variations,
\begin{equation}
U_j(\theta_j+\delta\theta_j)
=
U_j(\theta_j) e^{-i G_j \delta\theta_j},
\end{equation}
which gives
\begin{equation}
\widetilde{U}_j
=
U_j^\dagger(\theta_j)\, \widetilde{U}_{j+1}\, U_j(\theta_j)\, e^{-i G_j \delta\theta_j}.
\end{equation}
Since the construction is initialized at the identity and only small perturbations are inserted at each step, all intermediate operators $\widetilde{U}_j$ remain close to the identity throughout the iteration.
The MPO bond dimension required to represent an operator is governed by its operator entanglement. Because each $\widetilde{U}_j$ deviates only weakly from the identity, the operator entanglement remains small at every step, and the MPO admits a representation with relatively low bond dimension during the entire construction. In practice, this allows for efficient truncation after each iteration step without significantly compromising accuracy.
In particular, this procedure enables the efficient classical approximation of fidelities of the form $\left|\braket{\psi(\vtheta)}{\psi(\vtheta+\delta\vtheta)}\right|^2$ appearing in the \ac{SPSA} scheme introduced above.

\subsection{Why Stochastic Methods Fail in Practice}
\label{section:ComputingTheQGTAccurately}

Given the promising theoretical properties of stochastic methods such as \ac{SPSA}, a critical question is whether they can achieve the precision required for practical \ac{VarQTE} within reasonable computational budgets. 
To answer this, we analyze the \ac{SPSA} estimator under assumptions that are deliberately favorable to the method.
We ignore several inherent sources of error: the finite-difference error in approximating the Hessian, as well as any statistical, sampling, or truncation error in estimating the fidelity combination $\delta F$ in \cref{eq:approx_fidelity}. 
Instead, we replace $\delta F$ by its exact theoretical value,
\begin{equation}
  \delta F_{\text{exact}} = \spsapert^T g \spsapert,
\end{equation}
computed from a reference \ac{QGT} $g$. 
This gives an optimistic lower-bound estimate of the resources required to approximate the \ac{QGT} to a given precision. 
For additional context, we also compute a baseline approximation, $g_{\text{Cl}}$, obtained by rounding all circuit parameters to their nearest Clifford values. 
As evident from the structure of \cref{eq:qgt}, evaluating the \ac{QGT} at such Clifford points is classically efficient, making this a trivial but informative benchmark. 
Repeating this analysis for different circuits and parameter values did not change the qualitative conclusions.

We conduct the numerical investigation using a realistic quantum circuit. 
Specifically, we use the 63-qubit heavy-hex circuit ansatz presented in \cref{fig:scheduling}, which is later employed in the \ac{SKQD} experiments of Sec.~\ref{sec:SKQD}. 
We select the parameters from an arbitrary point along a \ac{VarQTE} trajectory. 
Using the method introduced in Sec.~\ref{sec:bowtiegradientandqgtcomputation}, we compute an exact reference value for the \ac{QGT}.

\begin{figure}[ht]
  \centering
  \includegraphics[width=0.5\linewidth]{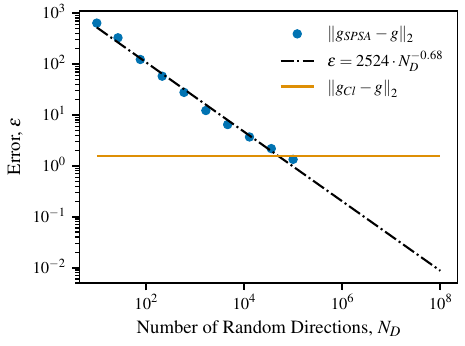}
  \caption{
    Comparison of the \ac{QGT} approximation error $\epsilon$ for the trivial Clifford approximation, shown in orange, and the \ac{SPSA} estimator, shown in blue. 
    A power-law fit to the \ac{SPSA} data, shown in black, is extrapolated beyond the number of random directions that can be computed directly.
  }
  \label{fig:QGT_approximation}
\end{figure}

The results are summarized in \cref{fig:QGT_approximation}, which shows the approximation error as a function of the number of random directions sampled in the \ac{SPSA} estimator. 
To match the error of the trivial Clifford approximation $g_{\text{Cl}}$, \ac{SPSA} already requires on the order of $10^5$ random directions. 
Achieving a more stringent error target, for example $\epsilon < 10^{-2}$, would require an estimated $10^8$ random directions, corresponding to the same order of distinct circuit evaluations. 
This is prohibitively expensive both for quantum-hardware implementations and for approximate classical simulators. 
Importantly, this estimate is obtained under highly favorable assumptions; including the omitted finite-difference and fidelity-estimation errors would only increase the required resources.

This analysis reveals a fundamental limitation of stochastic \ac{QGT} estimators in the regime relevant to this work. 
Although methods such as \ac{SPSA} have attractive scaling properties when measured only in terms of the number of parameters, the accuracy required for stable \ac{VarQTE} dynamics is too demanding. 
The ill-conditioned linear system in \cref{eq:sle} requires highly accurate \ac{QGT} estimates, and the number of stochastic samples needed to reach this accuracy is far beyond practical computational budgets. 
Thus, for the large-scale \ac{VarQTE} simulations considered here, the main obstacle is not merely the cost of evaluating individual fidelities, but the statistical cost of reconstructing a sufficiently accurate \ac{QGT} from random directions.

\section{Overlap Qubit Intersection}
\label{app:overlap_qubit_intersect}
Let $\tb{A}$ and $\tb{B}$ be two bowtie states acting on the qubit subsets $\mathcal{S}_A$ and $\mathcal{S}_B$, respectively.
Their inner product can be written as
\begin{equation}
\braket{\beta_A}{\beta_B} =
\sum_{\substack{
x\in\{0,1\}^{|\mathcal{S}_B\setminus\mathcal{S}_A|}\\[1mm]
y\in\{0,1\}^{|\mathcal{S}_A\setminus\mathcal{S}_B|}\\[1mm]
z\in\{0,1\}^{|\mathcal{S}_A\cap\mathcal{S}_B|}
}}
\beta^A_{x\times y\times z}\;
\beta^B_{x\times y\times z}.
\end{equation}
Here $x$, $y$, and $z$ denote bitstrings in the computational basis, and
$\beta^{A/B}_{x\times y\times z} = \braket{x\times y\times z}{\beta_A}$ denotes the corresponding amplitude of $\ket{\beta_{A/B}}$.

Since $\tb{A/B}$ only has support on $\mathcal{S}_{A/B}$, we have $\beta^{A/B}_{x\times y\times z}=0$ whenever $x/y\neq 0$.
These constraints can be encoded as:
\begin{equation}
\beta^{A}_{x\times y\times z} = \beta^{A}_{0\times y\times z}\,\delta_{x,0},
\qquad
\beta^{B}_{x\times y\times z} = \beta^{B}_{x\times 0\times z}\,\delta_{y,0}.
\end{equation}
Substituting the above into the overlap collapses the sums over $x$ and $y$ such that:
\begin{equation}
\braket{\beta_A}{\beta_B}
= \sum_{z\in\{0,1\}^{|\mathcal{S}_A \cap \mathcal{S}_B|}}
\beta^A_{0\times 0\times z}\;
\beta^B_{0\times 0\times z}.
\end{equation}
It follows that the computational cost is
$O\!\left(2^{|\mathcal{S}_A\cap \mathcal{S}_B|}\right)$,
rather than the naive
$O\!\left(2^{|\mathcal{S}_A\cup \mathcal{S}_B|}\right)$.
Moreover, with a precomputation stage that determines $\mathcal{S}_A\cap\mathcal{S}_B$ for each pair of states, each overlap reduces to a dot product over $2^{|\mathcal{S}_A\cap\mathcal{S}_B|}$ elements--which is highly efficient on classical hardware.

\section{Generalized QGT Formulation}
\label{appendix:aqc2mclachlan}

We begin by noting the structural similarity between the overlaps appearing in~\cref{eq:overlaps1,eq:overlaps2,eq:overlaps3,eq:overlaps4}. These expressions suggest a unified interpretation in which Hamiltonian terms act as infinitesimal generators of rotations appended to the variational circuit, thereby driving the time evolution directly in parameter space. This observation motivates the introduction of a \emph{generalized QGT}.
Consider an augmented parameter vector
\begin{equation}
    \tvtheta \coloneqq \left( \vtheta^{\mathsf T},\, i\,\Vec{c}^{\mathsf T} t \right)^{\mathsf T},
\end{equation}
where the additional components correspond to Hamiltonian terms weighted by their coefficients $\Vec{c}$. With this convention, the time derivative satisfies
\begin{equation}
    \partial_t R_{h_j}(i c_j t)\Bigr\rvert_{t = 0} = h_j,
\end{equation}
such that infinitesimal variations of these augmented parameters reproduce the action of the Hamiltonian. The parameters $\Vec{c} t$ therefore correspond exactly to those appearing in a first--order, depth--one Trotter circuit, which implements the exact dynamics in the limit $t \to 0$.
The QGT associated with this augmented parametrization defines the {generalized QGT}, denoted by $\Tilde{g}(\vtheta)$, which has the block structure
\begin{equation}
\Tilde{g}(\vtheta)=\Re
\begin{pmatrix}
  \begin{matrix}
   \text{QGT},\, g \\
   \braket{\beta_i}{\beta_j}
   -\braket{\beta_i}{\alpha}
   \braket{\alpha}{\beta_j}
  \end{matrix}
  & \vline &
  \begin{matrix}
   \text{Gradient},\, \Vec{b} \\
   \braket{\beta_i}{\gamma_j}
   -\braket{\beta_i}{\alpha}
   \braket{\alpha}{\gamma_j}
  \end{matrix}
  \\
\hline
  \begin{matrix}
   \text{Gradient},\, \Vec{b} \\
   \braket{\gamma_i}{\beta_j}
   -\braket{\gamma_i}{\alpha}
   \braket{\alpha}{\beta_j}
  \end{matrix}
  & \vline &
  \begin{matrix}
   \Var[H] \\
   \braket{\gamma_i}{\gamma_j}
   -\braket{\gamma_i}{\alpha}
   \braket{\alpha}{\gamma_j}
  \end{matrix}
\end{pmatrix}.
\end{equation}

This tensor naturally decomposes into three physically meaningful regions, from which the standard variational quantities can be recovered.

The upper--left block coincides with the standard QGT definition in Eq.~\cref{eq:qgt}. Explicitly, for $i,j \le N_p$,
\begin{equation}
\label{eq:genqgtQGT}
    g(\vtheta)_{i,j}
    =
    \Tilde{g}(\vtheta)_{i,j}
    =
    \Re\!\left(
    \braket{\beta_i}{\beta_j}
    -\braket{\beta_i}{\alpha}
    \braket{\alpha}{\beta_j}
    \right),
\end{equation}
where $N_p$ is the number of circuit parameters.
Furthermore, the real--time gradient can be recovered from the mixed block of $\Tilde{g}$ by contraction with the Hamiltonian coefficients:
\begin{align}
\label{eq:genqgtGrad}
    b_i^{\mathrm{real}}(\vtheta)
    &=
    \sum_{j>N_p} i c_j\, \Tilde{g}_{i,j}(\vtheta) \\
    &=
    \sum_{j>N_p} i c_j
    \Re\!\left(
    \braket{\beta_i}{\gamma_j}
    -\braket{\beta_i}{\alpha}
    \braket{\alpha}{\gamma_j}
    \right) \\
    &=
    \Im\!\left(
    \bra{\parvarpsi{i}} H \ket{\varpsi}
    - \bra{\parvarpsi{i}} \ket{\varpsi} E_{\vtheta}
    \right),
\end{align}
where we have used the definitions in
Eqs.~\cref{eq:alpha,eq:beta,eq:gamma}.
Interestingly, the lower--right block reproduces the energy variance as a quadratic form:
\begin{align}
\label{eq:genqgtVar}
    \Var[H]
    &=
    \sum_{i>N_p}\sum_{j>N_p} c_i c_j\, \Tilde{g}_{i,j}(\vtheta) \\
    &=
    \Re\!\left(
    \bra{\varpsi}H^2\ket{\varpsi}
    -\bra{\varpsi}H\ket{\varpsi}^2
    \right).
\end{align}

Finally, combining Eqs.~\cref{eq:genqgtQGT,eq:genqgtGrad,eq:genqgtVar}, the McLachlan error takes the compact form
\begin{equation}\label{eq:appendix_errorbound}
    \norm{e_t}^2_2
    =
    \Dot{\ttheta}^{\mathsf T} \Tilde{g} \Dot{\ttheta}
    =
    \Var[H]
    + \Dot{\vtheta}^{\mathsf T} g \Dot{\vtheta}
    -2 \Dot{\vtheta}^{\mathsf T} \Vec{b}.
\end{equation}
This expression matches the one derived in \cite{Zoufal_2023ErrorBounds} where they prove this quantity can be used to upper bound the infidelity between the target state and the variationally evolved state.

The generalized QGT also enables a compact constrained formulation of McLachlan's principle. Starting from
\begin{equation}
    g(\vtheta)\dvtheta = \Vec{b}(\vtheta),
\end{equation}
we may write
\begin{align}
    \dvtheta
    &=
    \underset{\Dot{\ttheta}}{\argmin}\;
    \Dot{\ttheta}^{\mathsf T} \Tilde{g} \Dot{\ttheta} \\
    &\text{s.t.} \quad
    \Dot{\Tilde{\theta}}_i = c_i \quad \forall i > N_p.
\end{align}
An equivalent penalty--based formulation is
\begin{equation}
    \dvtheta
    =
    \underset{\Dot{\ttheta}}{\argmin}\;
    \Bigl[
    \Var[H]
    + \Dot{\ttheta}^{\mathsf T} g \Dot{\ttheta}
    -2 \Dot{\ttheta}^{\mathsf T} \Vec{b}
    + \lambda \sum_{i>N_p} (\Dot{\Tilde{\theta}}_i - c_i)^2
    \Bigr].
\end{equation}

This formulation suggests possible extensions in which one allows effective Hamiltonian coefficients $\Vec{c}' \approx \Vec{c}$ to reduce the instantaneous variational error. Quantifying the potential accuracy gains of such approaches is beyond the scope of the present work.

\end{document}